\documentclass[11pt,a4paper]{article}
\usepackage{jheppub}

\usepackage[T1]{fontenc}
\usepackage{graphicx}
\usepackage{slashed}
\usepackage{amsmath,amsfonts,bbm,latexsym,amssymb}
\usepackage{epsfig}
\usepackage{array}

\usepackage{booktabs}
\usepackage[utf8]{inputenc}
\usepackage{fancyvrb}
\usepackage{subfig}
\newcommand{\newc}{\newcommand}
\newc{\be}{\begin{equation}}
\newc{\ee}{\end{equation}}
\newc{\bea}{\begin{eqnarray}}
\newc{\eea}{\end{eqnarray}}
\newc{\ol}{\overline}
\newc{\wt}{\widetilde}
\newc{\bs}{\boldsymbol}
\newc{\m}{\mathcal}
\newc{\la}{\langle}
\newc{\ra}{\rangle}

\newcommand{\beq}{\begin{eqnarray}}
\newcommand{\eeq}{\end{eqnarray}}
\newcommand{\bpmatrix}{\begin{pmatrix}}
\newcommand{\epmatrix}{\end{pmatrix}}

\renewcommand{\ol}{\text{1l}}

\renewcommand{\Re}{\text{Re}\!}

\renewcommand{\eqref}[1]{eq.~(\ref{#1})}

\newcommand{\bc}{\begin{center}}
\newcommand{\ec}{\end{center}}

\newcommand{\s}{\newline \vspace*{-3.5mm}}

\def\be{\begin{equation}}
\def\ee{\end{equation}}
\def\ba{\begin{eqnarray}}
\def\ea{\end{eqnarray}}
\def\ra{\rightarrow}
\def\vb#1{\vbox to #1 pt{}}

\def\m#1{m_{#1}}

\def\ltap{\;\centeron{\raise.35ex\hbox{$<$}}{\lower.65ex\hbox{$\sim$}}\;}
\def\gtap{\;\centeron{\raise.35ex\hbox{$>$}}{\lower.65ex\hbox{$\sim$}}\;}
\def\vb#1{\vbox to #1 pt{}}

\begin{document}
\begin{flushright}
CFTP/17-008\\[-1mm]
DESY 17-207\\[-1mm]
KA-TP-40-2017\\[-1mm]
\end{flushright}
\vspace*{1cm}

\title{The C2HDM revisited}

\author[a]{Duarte Fontes,}
\author[b]{Margarete M\"{u}hlleitner,}
\author[a]{Jorge C. Rom\~ao,}
\author[c,d,e]{Rui Santos,}
\author[a]{Jo\~{a}o P. Silva}
\author[b,f]{and Jonas Wittbrodt}

\affiliation[a]{Departamento de F\'{\i}sica and CFTP,
Instituto Superior T\'{e}cnico, Universidade de Lisboa, \\
Avenida Rovisco Pais 1, 1049-001 Lisboa, Portugal}
\affiliation[b]{Institute for Theoretical Physics,
Karlsruhe Institute of Technology, \\
76128 Karlsruhe, Germany}
\affiliation[c]{ISEL - Instituto Superior de Engenharia de Lisboa, \\
Instituto Polit\'ecnico de Lisboa 1959-007 Lisboa, Portugal}
\affiliation[d]{Centro de F\'{\i}sica Te\'{o}rica e Computacional, Faculdade de Ci\^{e}ncias, \\
Universidade de Lisboa, Campo Grande, Edif\'{\i}cio C8,
1749-016 Lisboa, Portugal}
\affiliation[e]{LIP, Departamento de F\'{\i}sica,
Universidade do Minho, 4710-057 Braga, Portugal}
\affiliation[f]{Deutsches Elektronen-Synchrotron DESY, Notkestra{\ss}e 85, D-22607
Hamburg, Germany}
\emailAdd{duartefontes@tecnico.ulisboa.pt}
\emailAdd{margarete.muehlleitner@kit.edu}
\emailAdd{jorge.romao@tecnico.ulisboa.pt}
\emailAdd{rasantos@fc.ul.pt}
\emailAdd{jpsilva@cftp.ist.utl.pt}
\emailAdd{jonas.wittbrodt@desy.de}

\date{\today}

\abstract{The complex two-Higgs doublet model is one of the simplest ways to extend the scalar sector of the Standard Model to include a new source of CP-violation.
The model has been used as a benchmark model to search for CP-violation at the LHC and as a possible explanation for the matter-antimatter asymmetry of the Universe.
In this work, we re-analyse in full detail the softly broken $\mathbb{Z}_2$
symmetric complex two-Higgs doublet model (C2HDM).
We provide the code {\tt C2HDM\_HDECAY} implementing the C2HDM in the well-known {\tt HDECAY} program which calculates the decay widths including the state-of-the-art higher order QCD corrections and the relevant off-shell decays.
Using {\tt C2HDM\_HDECAY} together with the most relevant theoretical and experimental constraints, including electric dipole moments (EDMs), we review the parameter space of the model and discuss its phenomenology. In particular, we find cases where large CP-odd couplings to fermions are still allowed and provide benchmark points for these scenarios. We examine the prospects of discovering CP-violation at the LHC and show how theoretically motivated measures of CP-violation correlate with observables.}



\maketitle

\section{Introduction}
\label{sec:intro}

After the discovery of the Higgs boson by the ATLAS~\cite{Aad:2012tfa}
and CMS~\cite{Chatrchyan:2012ufa} collaborations the Large Hadron
Collider (LHC) has now entered the Run 2 stage with a 13 TeV centre-of-mass energy.
The search for physics beyond the Standard Model (BSM) is now the main goal of the LHC experiments.
An important motivation of BSM physics is to provide new sources of CP-violation to fulfil the three Sakharov criteria for baryogenesis~\cite{Sakharov:1967dj}.
The simplest extension of the scalar sector that can provide a new source of CP-violation,
the two-Higgs doublet model (2HDM) \cite{Lee:1973iz}, can address this issue.
Reviews of the 2HDM can be found in~\cite{Gunion:1989we, Branco:2011iw, Ivanov:2017dad}.

One of the simplest ways of extending the SM with a CP-violating scalar sector without the addition of new fermions is to add a scalar doublet to the SM content and build a 2HDM softly broken $\mathbb{Z}_2$ symmetric potential with two complex parameters.
This model, now known as C2HDM, was first discussed in~\cite{Ginzburg:2002wt} and has only one independent CP-phase and a simple limit leading to its CP-conserving version.
The model has been the subject of many studies~\cite{Khater:2003wq, ElKaffas:2007rq, Grzadkowski:2009iz, Arhrib:2010ju, Barroso:2012wz, Inoue:2014nva,Cheung:2014oaa, Fontes:2014xva, Fontes:2015mea, Chen:2015gaa}.
More recently a comparison with a number of other extensions of the SM was performed~\cite{Muhlleitner:2017dkd}.
The NLO QCD corrections to double Higgs decays were also recently calculated in~\cite{Grober:2017gut}.

One of the primary goals of the LHC Run 2 is to look for new particles but also to probe the CP parity of both the discovered Higgs boson and of any further scalars yet undiscovered.
Due to its simplicity, the C2HDM is the ideal benchmark model to test the CP quantum numbers of the scalars at the LHC.
The CP-nature of the scalars, in their Yukawa couplings, can be probed directly either in the production or the decays of the Higgs bosons.
In this case,
it is the relation between the scalar and the pseudoscalar components in the Yukawa couplings that is probed.
Some examples of the use of asymmetries to probe the CP-nature of the
Higgs boson in the top Yukawa coupling were discussed in~\cite{Gunion:1996xu, Boudjema:2015nda, AmorDosSantos:2017ayi}
while the decays of the tau leptons were used to probe the tau Yukawa coupling~\cite{Berge:2008wi, Berge:2008dr, Berge:2011ij, Berge:2014sra, Berge:2015nua}.
Correlations in the momentum distributions of leptons produced in the
decays of the Higgs boson to gauge bosons were used to probe the CP-nature of the Higgs boson couplings to gauge bosons~\cite{Choi:2002jk, Buszello:2002uu, Godbole:2007cn} by the ATLAS \cite{Aad:2015mxa} and CMS \cite{Khachatryan:2016tnr} collaborations.
The most general CP-violating $HVV$ coupling was used, and limits were set on the anomalous couplings.
However, the C2HDM has SM-like couplings to the gauge bosons - it is just the SM coupling multiplied by a number.
The anomalous couplings only appear at loop level and are consequently very small.
Therefore, in this model, only the Yukawa couplings can lead to direct observations of CP-violation.

There are, however, other ways to probe CP-violation using only inclusive observables.
As proposed in \cite{Branco:1999fs, Fontes:2015xva} several combinations of three simultaneously observed Higgs decay modes can constitute an undoubtable sign of CP-violation.
In a CP-conserving model, a decay of the type $H_i \rightarrow H_j Z$ would imply opposite CP parities for $H_i$ and $H_j$.
In a renormalisable theory, a Higgs boson decaying to a pair of gauge bosons has to be CP-even\footnote{A CP-odd scalar decays to a pair of gauge bosons at one-loop.
It was shown for the CP-conserving 2HDM that the corresponding width is several orders of magnitude smaller than the corresponding tree-level one, $H \to ZZ$~\cite{Arhrib:2006rx, Bernreuther:2010uw}.}.

Hence, the combination of the decays $H_i \rightarrow H_j Z$, $H_i \rightarrow Z Z$ and $H_j \rightarrow Z Z$ is a clear sign of CP-violation.
In \cite{Fontes:2015xva} seven classes of decays were defined, some of which signal CP-violation for any extension of the SM, while others are not possible in a CP-conserving 2HDM but can occur in the C2HDM or even in models with 3 CP-even states.
An example of the latter is $H_i \to ZZ$ (with $i=1,2,3$) that could signal CP-violation but could also happen in a model with at least 3 CP-even states.
In some extensions of the SM, even the ones with just two extra scalars, as is the case of the SM plus a complex singlet, all $H_i$ are CP-even, and therefore the decays $H_i \to ZZ$ all happen at tree-level.
From the above classes, we are especially interested in the ones that include the already observed decay of the SM-like Higgs boson, denoted by $h_{125}$, to gauge bosons, $h_{125} \to ZZ$.
Furthermore, decays of the type $H_i \to ZZ$ and $H_i \to H_j Z$ were the subject of searches during Run 1 which will proceed during Run 2.
Therefore, if a new Higgs boson is observed in the final states with gauge bosons and also in the final state $h_{125} Z$ the scalar sector is immediately established to be CP-violating.

In the C2HDM there is only one independent CP-violating phase.
Hence, the only three basis-invariant quantities that signal CP-violation \cite{Lavoura:1994fv, Botella:1994cs} are all related and proportional to that phase.
 In this work, we are interested in finding a relation between the production rates in each of the CP-violating classes and variables that signal CP-violation.
Therefore we will test a number of variables in the most interesting classes that include the already observed $h_{125} \to ZZ$.
Finally, decays of the type $H_j \to H_i h_{125}$ can only happen in very specific models and searches for this particular channel are important and should be a priority for Run 2.

The LHC Run 2 will bring increasing precision in the measurement of Higgs production and decay rates.
The ATLAS and CMS collaborations have started testing new models such as the singlet extension of the SM and the CP-conserving 2HDM.
To probe other extensions of the SM new tools are needed to increase the precision both in the Higgs production rates and also in the Higgs decay widths.
Hence, one of the main purposes of this work is the release of the {\tt C2HDM\_HDECAY} code, an implementation of the C2HDM in {\tt HDECAY v6.51}~\cite{Djouadi:1997yw, Butterworth:2010ym}.
The code is entirely self-contained, and the widths include the most important state-of-the-art higher order QCD corrections and the relevant off-shell decays.

The paper is organised as follows.
In section~\ref{sec:model} we describe the model and in section~ \ref{sec:c2hdmspace} we discuss the parameter space of the model given the most relevant theoretical and up to date experimental constraints.
We identify situations where $h_{125}$ could have remarkable properties and give benchmark points for these scenarios.
Namely, situations where its couplings to fermions have a large CP-odd component and the possibilities that $h_{125}$ is the heaviest of the three scalars in the C2HDM.
In section~\ref{sec:measure}, we discuss the production rates of processes that constitute classes of CP-violating decays and relate them with a number of variables that can probe CP-violation.
In section \ref{sec:higgstohiggs} we discuss Higgs-to-Higgs decays in the framework of the C2HDM.
Our conclusions are presented in section~\ref{sec:concl}.
In appendix~\ref{app:c2hdm_FR} we write the Feynman rules for the C2HDM in the unitary gauge.
In appendix~\ref{app:c2hdmhdecay} we describe the  {\tt C2HDM\_HDECAY} code.

\section{The complex two-Higgs doublet model}
\label{sec:model}
The version of the complex two-Higgs doublet model we discuss in this work has an explicitly
CP-violating scalar potential, with a  softly broken $\mathbb{Z}_2$ symmetry $\Phi_1 \ra \Phi_1, \Phi_2 \ra -\Phi_2$
written as
\beq
V &=& m_{11}^2 |\Phi_1|^2 + m_{22}^2 |\Phi_2|^2
- \left(m_{12}^2 \, \Phi_1^\dagger \Phi_2 + h.c.\right)
+ \frac{\lambda_1}{2} (\Phi_1^\dagger \Phi_1)^2 +
\frac{\lambda_2}{2} (\Phi_2^\dagger \Phi_2)^2 \nonumber \\
&& + \lambda_3
(\Phi_1^\dagger \Phi_1) (\Phi_2^\dagger \Phi_2) + \lambda_4
(\Phi_1^\dagger \Phi_2) (\Phi_2^\dagger \Phi_1) +
\left[\frac{\lambda_5}{2} (\Phi_1^\dagger \Phi_2)^2 + h.c.\right] \;.
\eeq
Due to the hermiticity of the Lagrangian,
all couplings are real except for $m_{12}^2$ and $\lambda_5$.
We write each of the doublets $\Phi_i$ $(i=1,2)$ as an expansion around
the real vacuum expectation values (VEVs) $v_{1}$ and $v_{2}$,
in terms of the charged complex fields ($\phi_i^+$)
and the real neutral fields ($\rho_i$ and $\eta_i$).
The doublets then read
\beq
\Phi_1 = \left(
\begin{array}{c}
\phi_1^+ \\
\frac{v_1 + \rho_1 + i \eta_1}{\sqrt{2}}
\end{array}
\right) \qquad \mbox{and} \qquad
\Phi_2 = \left(
\begin{array}{c}
\phi_2^+ \\
\frac{v_2 + \rho_2 + i \eta_2}{\sqrt{2}}
\end{array}
\right) \;. \label{eq:2hdmdoubletexpansion}
\eeq
The minimum conditions for the potential are
\beq
m_{11}^2 v_1 + \frac{\lambda_1}{2} v_1^3 + \frac{\lambda_{345}}{2} v_1
v_2^2 &=& \textrm{Re} \left(m_{12}^2\right) v_2 \;, \label{eq:mincond1} \\
m_{22}^2 v_2 + \frac{\lambda_2}{2} v_2^3 + \frac{\lambda_{345}}{2} v_1^2
v_2 &=& \textrm{Re} \left(m_{12}^2\right) v_1 \;, \label{eq:mincond2} \\
2\, \mbox{Im} (m_{12}^2) &=& v_1 v_2 \mbox{Im} (\lambda_5)
\;, \label{eq:mincond3}
\eeq
where we have introduced
\beq
\lambda_{345} \equiv \lambda_3 + \lambda_4 + \mbox{Re} (\lambda_5) \;.
\eeq
We define two CP-violating phases $\phi (m_{12}^2)$ and $\phi (\lambda_5)$ as
\begin{equation}
m_{12}^2 = |m_{12}^2|\, e^{i\, \phi (m_{12}^2)}\, ,
\qquad \qquad \lambda_5 = |\lambda_5|\, e^{i\, \phi (\lambda_5)} \, .
\end{equation}
Equation~(\ref{eq:mincond3}) shows us that these two
phases are not independent.
We can re-write eq.~(\ref{eq:mincond3}) as
\begin{equation}
2 \Re{\, (m_{12}^2)} \, \tan \phi (m_{12}^2) = v_1 v_2 \, \Re{\, (\lambda_5)} \, \tan \phi (\lambda_5) \, .
\end{equation}
We choose both vacuum expectation values $v_1$ and $v_2$ to be real
which together with the condition
$\phi(\lambda_5) \neq 2\, \phi(m_{12}^2)$~\cite{Ginzburg:2002wt}
ensures that the two phases cannot be removed simultaneously.
Otherwise we are in the CP-conserving limit of the model.

The Higgs basis \cite{Lavoura:1994fv,Botella:1994cs}
$\{ {\cal H}_1, {\cal H}_2 \}$ is defined by the rotation
\beq
\left( \begin{array}{c} {\cal H}_1 \\ {\cal H}_2 \end{array} \right) =
R^T_H \left( \begin{array}{c} \Phi_1 \\
    \Phi_2 \end{array} \right) \equiv
\left( \begin{array}{cc} c_\beta & s_\beta \\ - s_\beta &
    c_\beta \end{array} \right) \left( \begin{array}{c} \Phi_1 \\
    \Phi_2 \end{array} \right) \;,
\eeq
with
\beq
\tan \beta \equiv \frac{v_2}{v_1} \;,
\eeq
and the doublets in the Higgs basis are written as
\beq
{\cal H}_1 = \left( \begin{array}{c} G^\pm \\ \frac{1}{\sqrt{2}} (v + H^0
    + i G^0) \end{array} \right) \quad \mbox{and} \qquad
{\cal H}_2 = \left( \begin{array}{c} H^\pm \\ \frac{1}{\sqrt{2}} (R_2
    + i I_2) \end{array} \right) \;.
\eeq
The SM VEV $v = \sqrt{v_1^2 + v_2^2}$ along with the Goldstone bosons
$G^\pm$ and $G^0$ is now in ${\cal H}_1$, while the charged Higgs mass
eigenstates $H^\pm$ are in ${\cal H}_2$. The neutral Higgs
mass eigenstates $H_i$ ($i=1,2,3$) are obtained from the neutral
components of the C2HDM basis, $\rho_1$, $\rho_2$ and $\rho_3 \equiv I_2$,
via the rotation
\beq
\left( \begin{array}{c} H_1 \\ H_2 \\ H_3 \end{array} \right) = R
\left( \begin{array}{c} \rho_1 \\ \rho_2 \\ \rho_3 \end{array} \right)
\;.
\label{eq:c2hdmrot}
\eeq

The model has three neutral particles with no definite CP quantum numbers,
$H_1$, $H_2$ and $H_3$, and two charged scalars $H^\pm$. The mass matrix of the neutral scalar states
\beq
({\cal M}^2)_{ij} = \left\langle \frac{\partial^2 V}{\partial \rho_i
  \partial \rho_j} \right\rangle \;,
\label{eq:c2hdmmassmat}
\eeq
is diagonalised via the orthogonal matrix $R$~\cite{ElKaffas:2007rq}.
That is,
\beq
R {\cal M}^2 R^T = \mbox{diag} (m_{H_1}^2, m_{H_2}^2, m_{H_3}^2) \;,
\eeq
for which we choose the form
\be
R =
\left(
\begin{array}{ccc}
c_1 c_2 & s_1 c_2 & s_2\\
-(c_1 s_2 s_3 + s_1 c_3) & c_1 c_3 - s_1 s_2 s_3  & c_2 s_3\\
- c_1 s_2 c_3 + s_1 s_3 & -(c_1 s_3 + s_1 s_2 c_3) & c_2 c_3
\end{array}
\right)\, ,
\label{matrixR}
\ee
with $s_i = \sin{\alpha_i}$,
$c_i = \cos{\alpha_i}$ ($i = 1, 2, 3$),
and
\be
- \pi/2 < \alpha_1 \leq \pi/2,
\hspace{5ex}
- \pi/2 < \alpha_2 \leq \pi/2,
\hspace{5ex}
- \pi/2 < \alpha_3 \leq \pi/2.
\label{range_alpha}
\ee
The Higgs boson masses are ordered such that $m_{H_1} \le m_{H_2} \le m_{H_3}$.

Note that the mass basis and the Higgs basis are related through
\beq
\left( \begin{array}{c} H_1 \\ H_2 \\ H_3 \end{array} \right) = T^T
\left( \begin{array}{c} H^0 \\ R_2 \\ I_2 \end{array} \right) \;,
\eeq
where the matrix $T$ used in ref.~\cite{Branco:2011iw}
for the expression of the oblique radiative corrections
is defined as
\beq
T^T = R \widetilde{R}_H
\eeq
and $\widetilde{R}_H$ is given by
\beq
\widetilde{R}_H = \left( \begin{array}{cc} R_H & 0 \\ 0 & 1 \end{array}
\right) \;.
\eeq

Our choice of the 9 independent parameters of the C2HDM is:
$v = \sqrt{v_1^2 + v_2^2}$, $\tan \beta$, $m_{H^\pm}$,
$\alpha_1$, $\alpha_2$, $\alpha_3$, $m_{H_1}$, $m_{H_2}$, and $\textrm{Re}(m_{12}^2)$.
With this choice, the mass of the heaviest neutral scalar is a dependent parameter,
given by
\be
m_{H_3}^2 = \frac{m_{H_1}^2\, R_{13} (R_{12} \tan{\beta} - R_{11})
+ m_{H_2}^2\ R_{23} (R_{22} \tan{\beta} - R_{21})}{R_{33} (R_{31} - R_{32} \tan{\beta})},
\label{m3_derived}
\ee
and the parameter space points will have to comply with $m_{H_3} > m_{H_2}$.

We will briefly describe here the couplings of the Higgs bosons
with the remaining SM fields.
A longer list is provided in appendix~\ref{app:c2hdm_FR},
and the full set is contained in the web page~\cite{C2HDM_FR}.
The Higgs couplings to the massive gauge bosons $V=W,Z$ are given by
\beq
i \, g_{\mu\nu} \, c(H_i VV) \, g_{H^{\text SM} VV} \;, \label{eq:gaugecoupdef}
\eeq
where $g_{H^{\text SM} VV}$ denotes the corresponding SM Higgs coupling,
given by
\beq
g_{H^{\text SM} VV} = \left\{\begin{array}{c} g M_W \qquad \qquad \quad V=W
\\ g M_Z /\cos\theta_W \qquad V=Z \end{array} \right. \qquad
\eeq
where $g$ is the $SU(2)$ gauge coupling and $\theta_W$ is the Weinberg angle.
The effective couplings can be written as
\beq
c(H_i VV) = T_{1i} = c_\beta R_{i1} + s_\beta R_{i2} \;. \label{eq:c2dhmgaugecoup}
\eeq
The Yukawa sector is built by extending the $\mathbb{Z}_2$ symmetry to the fermion
fields such that flavour changing neutral currents (FCNC) are absent at
tree-level~\cite{Glashow:1976nt,Paschos:1976ay}.
There are four possible $\mathbb{Z}_2$ charge assignments and therefore four
different types of 2HDMs described in table~\ref{tab:types}.
\begin{table}
\begin{center}
\begin{tabular}{rccc} \toprule
& $u$-type & $d$-type & leptons \\ \midrule
Type I & $\Phi_2$ & $\Phi_2$ & $\Phi_2$ \\
Type II & $\Phi_2$ & $\Phi_1$ & $\Phi_1$ \\
Lepton-Specific & $\Phi_2$ & $\Phi_2$ & $\Phi_1$ \\
Flipped & $\Phi_2$ & $\Phi_1$ & $\Phi_2$ \\ \bottomrule
\end{tabular}
\caption{The four Yukawa types of the softly broken $\mathbb{Z}_2$-symmetric 2HDM. \label{tab:types}}
\end{center}
\end{table}
The Yukawa Lagrangian has the form
\beq
{\cal L}_Y = - \sum_{i=1}^3 \frac{m_f}{v} \bar{\psi}_f \left[ c^e(H_i
  ff) + i c^o(H_i ff) \gamma_5 \right] \psi_f H_i \;, \label{eq:yuklag}
\eeq
where $\psi_f$ denote the fermion fields with mass $m_f$. The
coefficients of the CP-even and of the CP-odd part of the Yukawa
coupling, $c^e(H_i ff)$ and $c^o (H_i ff)$, are presented in table~\ref{tab:yukcoup}.
\begin{table}
\begin{center}
\begin{tabular}{rccc} \toprule
& $u$-type & $d$-type & leptons \\ \midrule
Type I & $\frac{R_{i2}}{s_\beta} - i \frac{R_{i3}}{t_\beta} \gamma_5$
& $\frac{R_{i2}}{s_\beta} + i \frac{R_{i3}}{t_\beta} \gamma_5$ &
$\frac{R_{i2}}{s_\beta} + i \frac{R_{i3}}{t_\beta} \gamma_5$ \\
Type II & $\frac{R_{i2}}{s_\beta} - i \frac{R_{i3}}{t_\beta} \gamma_5$
& $\frac{R_{i1}}{c_\beta} - i t_\beta R_{i3} \gamma_5$ &
$\frac{R_{i1}}{c_\beta} - i t_\beta R_{i3} \gamma_5$ \\
Lepton-Specific & $\frac{R_{i2}}{s_\beta} - i \frac{R_{i3}}{t_\beta} \gamma_5$
& $\frac{R_{i2}}{s_\beta} + i \frac{R_{i3}}{t_\beta} \gamma_5$ &
$\frac{R_{i1}}{c_\beta} - i t_\beta R_{i3} \gamma_5$ \\
Flipped & $\frac{R_{i2}}{s_\beta} - i \frac{R_{i3}}{t_\beta} \gamma_5$
& $\frac{R_{i1}}{c_\beta} - i t_\beta R_{i3} \gamma_5$ &
$\frac{R_{i2}}{s_\beta} + i \frac{R_{i3}}{t_\beta} \gamma_5$ \\ \bottomrule
\end{tabular}
\caption{Yukawa couplings of the Higgs
  bosons $H_i$ in the C2HDM, divided by the corresponding SM Higgs couplings. The expressions correspond to
  $[c^e(H_i ff) +i c^o (H_i ff) \gamma_5]$ from
  eq.~(\ref{eq:yuklag}). \label{tab:yukcoup}}
\end{center}
\end{table}

These couplings were implemented in the code {\tt HDECAY}~\cite{Djouadi:1997yw,Butterworth:2010ym} which provides
all Higgs decay widths and branching ratios including the state-of-the-art higher
order QCD corrections and off-shell decays. The description of the new  {\tt C2HDM\_HDECAY} code is presented
in appendix~\ref{app:c2hdmhdecay}.

\section{The C2HDM Parameter Space \label{sec:c2hdmspace}}

\subsection{Experimental and Theoretical Restrictions}

The C2HDM was implemented as a model class in
{\tt ScannerS}~\cite{Coimbra:2013qq,ScannerS}.
The most relevant theoretical and experimental bounds are either built
in the code or acessible via interfaces with other codes. We have
imposed all available constraints on the model and performed a parameter scan. The resulting viable points are the basis for our phenomenological analyses for the LHC Run 2.

The theoretical bounds included in {\tt ScannerS} are boundness from below and
perturbative unitarity~\cite{Kanemura:1993hm, Akeroyd:2000wc,Ginzburg:2003fe}.
Contrary to the SM, in the 2HDM coexisting minima can occur at tree-level. Therefore we also force the minimum to be global~\cite{Ivanov:2015nea}, precluding the possibility of vacuum decay. The points generated comply with electroweak precision measurements, making
use of the oblique parameters $S$, $T$ and $U$~\cite{Branco:2011iw}. We ask for a
$2\sigma$ compatibility of $S$, $T$ and $U$ with the SM fit presented in~\cite{Baak:2014ora}.
The full correlation among these parameters is taken into account.

The charged sector of the C2HDM has exactly the same couplings as in the
2HDM.\footnote{When mentioning simply the 2HDM,
we will be referring to the CP-conserving, softly broken $\mathbb{Z}_2$ symmetric 2HDM.}
Therefore, the exclusion bounds on
the $m_{H^\pm}-t_\beta$ plane can be imported from the 2HDM.
The most constraining bounds on this plane come from the measurements of $B \to X_s \gamma$ \cite{Deschamps:2009rh,Mahmoudi:2009zx,Hermann:2012fc,Misiak:2015xwa,Misiak:2017bgg}. The latest $2\sigma$ bounds on this
plane were obtained in~\cite{Misiak:2017bgg} and force the charged Higgs mass to be
$m_{H^\pm} > 580 \mbox{ GeV}$ for models Type II and Flipped,
almost independently of $\tan \beta$.
Due to the structure of the charged Higgs couplings to fermions, in models Type I and Lepton-Specific
the bound has a strong dependence on $\tan \beta$. In fact, for $\tan \beta \approx 1$ the bound is about
$400  \mbox{ GeV}$ while the LEP bound derived from $e^+ e^- \to H^+ H^-$~\cite{Abbiendi:2013hk}
(approximately $100  \mbox{ GeV}$) is recovered for $\tan \beta \approx 1.8$.
We further apply the flavour constraints from $R_b$~\cite{Haber:1999zh,Deschamps:2009rh}.
All the constraints are checked as $2\sigma$ exclusion bounds on the $m_{H^\pm}-t_\beta$ plane.

The SM-like Higgs boson is denoted by $h_{125}$ and has a mass of
$m_{h_{125}} = 125.09 \; \mbox{GeV}$ \cite{Aad:2015zhl}. We exclude points of the parameter space with
the discovered Higgs signal built by two nearly degenerate Higgs boson states
by forcing the non-SM scalar masses to be outside the mass window
$m_{h_{125}} \pm 5$~GeV. Compatibility with the exclusion bounds from Higgs searches is checked with the
{\tt HiggsBounds} code \cite{Bechtle:2008jh,Bechtle:2011sb,Bechtle:2013wla},
while the individual signal strengths for the SM-like
Higgs boson are forced to be within $2 \sigma$ of the fits presented in \cite{Khachatryan:2016vau}.
Branching ratios and decay widths of all Higgs bosons are calculated with the
{\tt C2HDM\_HDECAY} code, described in appendix~\ref{app:c2hdmhdecay}, which is an implementation of the
C2HDM model into {\tt HDECAY v6.51}. The code has state-of-the-art QCD corrections and
off-shell decays, but off-shell decays of one scalar into two are not included. The Higgs boson production cross sections
 via gluon fusion ($ggF$) and $b$-quark
fusion ($bbF$) are calculated with {\tt SusHiv1.6.0} \cite{Harlander:2012pb,Harlander:2016hcx},
which is interfaced with {\tt ScannerS}, at next-to-next-to-leading
order (NNLO) QCD. As the neutral scalars have no definite CP,
we need to combine the CP-odd and the CP-even contributions by
summing them incoherently. That is,
$\mu_F$ is given by
\beq
\mu_F = \frac{\sigma^{\text{even}}_{\text{C2HDM}} (ggF)
  +\sigma^{\text{even}}_{\text{C2HDM}} (bbF)
  +\sigma^{\text{odd}}_{\text{C2HDM}} (ggF) +
  \sigma^{\text{odd}}_{\text{C2HDM}}
  (bbF)}{\sigma^{\text{even}}_{\text{SM}} (ggF)} \;,
\eeq
where in the denominator we neglected the $bbF$ cross section since in the SM it is much
smaller than gluon fusion production.
The CP-even Higgs production cross sections in association with a vector boson  ($VH$)
and in vector boson fusion ($VBF$) give rise to the normalised production strengths
\beq
\mu_V = \frac{\sigma^{\text{even}}_{\text{C2HDM}}
  (VBF)}{\sigma^{\text{even}}_{\text{SM}} (VBF)} =
\frac{\sigma^{\text{even}}_{\text{C2HDM}}
  (VH)}{\sigma^{\text{even}}_{\text{SM}} (VH)} = c^2 (H_i VV) \;,
\eeq
because QCD corrections cancel at NLO. The effective couplings are defined in eq.~(\ref{eq:gaugecoupdef}).

Both CP-even and CP-odd components
contribute to the cross sections for associated production with fermions. As these have different QCD corrections  \cite{Hafliger:2005aj},
we opted for using leading order production cross sections, and write the strengths as
\beq
\mu_{\text{assoc}} = \frac{\sigma_{\text{C2HDM}}
  (ffH_i)}{\sigma_{\text{SM}} (ffH)} = c^e (H_i ff)^2 + c^o (H_i ff)^2 \;,
\eeq
with the coupling coefficients defined in eq.~(\ref{eq:yuklag}).
We then use  {\tt HiggsBounds} via the {\tt ScannerS}
interface to ensure compatibility at $2\sigma$ with all available collider data.
Regarding the SM-like Higgs boson, the parameter space is forced
to be in agreement with the fit results from \cite{Khachatryan:2016vau}.
That is, the values
\beq
\frac{\mu_F}{\mu_V} \;, \quad \mu_{\gamma\gamma} \;, \quad
\mu_{ZZ} \;, \quad \mu_{WW} \;, \quad  \mu_{\tau\tau} \;, \quad
\mu_{bb} \;,
\label{exp_constraints}
\eeq
given in \cite{Khachatryan:2016vau}, with $\mu_{xx}$ defined as
\beq
\mu_{xx} = \mu_F \, \frac{\mbox{BR}_{\text{C2HDM}} (H_i \to
  xx)}{\mbox{BR}_{\text{SM}} (H^{SM} \to xx)}
\eeq
for $H_i \equiv h_{125}$, are within $2 \sigma$ of the fitted
experimental values. Here $H^{SM}$ denotes the SM
  Higgs boson with a mass of 125.09~GeV.
As the C2HDM preserves custodial symmetry, $\mu_{ZZ} = \mu_{WW} \equiv \mu_{VV}$,
we can combine the lower $2 \sigma$ bound from
$\mu_{ZZ}$ with the upper bound on $\mu_{WW}$ \cite{Khachatryan:2016vau}, meaning
\beq
0.79 < \mu_{VV} < 1.48 \;.
\eeq
We use this method for simplicity. Note that performing a fit to current Higgs data is likely to give a stronger bound than this approach.

The C2HDM is a model with explicit CP-violation in the scalar sector.
Therefore, there are a number of experiments that allow us to constrain
the amount of CP-violation in the model.
The most restrictive bounds on the CP-phase~\cite{Inoue:2014nva}
(see also
\cite{Buras:2010zm, Cline:2011mm, Jung:2013hka, Shu:2013uua,
  Brod:2013cka,Basler:2017uxn})
originate from the
ACME~\cite{Baron:2013eja} results on the ThO molecule electric dipole moment (EDM).
All points in parameter
space have to conform to the ACME experimental results. As we will discuss in detail
later, the bounds can only be evaded either in the CP-conserving limit of the model
or in scenarios where cancellations between diagrams with different neutral scalar
particles occur~\cite{Jung:2013hka, Shu:2013uua}.
The cancellations are related to the orthogonality of the $R$ matrix in the case of almost degenerate
scalars~\cite{Fontes:2014xva}.
Our results are required to be compatible with the measured EDM values in
\cite{Baron:2013eja} at 90\% C.L.
We finalise with a word of caution
regarding the electron EDM. The authors of ref.~\cite{Jung:2013hka}
argue that the EDM could be up to one order of magnitude larger than the bound
presented by ACME, due to the large uncertainties in its extraction from the
experimental data.


In our scan, one of the Higgs bosons $H_i$ is identified with $h_{125}$.
One of the other neutral Higgs bosons is varied between $30\mathrm{\;
  GeV}\leq m_{H_i}<1\mathrm{\; TeV}$
while the third neutral Higgs boson is not an independent parameter and is calculated
by {\tt ScannerS}, but its mass is forced to be in the same interval.
In Type II and Flipped, the charged Higgs boson mass is forced to be in the range
\beq
580 \mbox{ GeV } \le m_{H^\pm} < 1 \mbox{ TeV } \;,
\eeq
while in Type I and Lepton-Specific we choose
\beq
80 \mbox{ GeV } \le m_{H^\pm} < 1 \mbox{ TeV } \;.
\eeq
Taking into account all the constraints, in order to optimise
the scan, we have chosen the following regions for the remaining input parameters:
$0.8 \le \tan \beta  \le 35$, $- \frac{\pi}{2} \le \alpha_{1,2,3} < \frac{\pi}{2}$ and
$ 0 \mbox{ GeV}^2 \le \mbox{Re}(m_{12}^2)  < 500 000 \mbox{ GeV}^2$~\footnote{We have generated a sample with
just the theoretical constraints and the number of points with a negative $\mbox{Re}(m_{12}^2)$
is of the order of 1 in 10 million. When we further impose the experimental constraints all such
points vanish.}.


\subsection{\label{subsec:constraints}Constraints on the Parameter
Space: $c^e$ versus $c^o$}

In this subsection, we confront the C2HDM parameter space with all the restrictions presented above.
Our aim is to see what is the structure of the remaining parameter space and, in particular, to study the CP-nature of the 125 GeV scalar, encoded in the couplings $c_f^e \equiv c^e(h_{125} f f)$ and $c_f^o \equiv c^o(h_{125}ff)$
of eq.~(\ref{eq:yuklag}).
These test the CP content of $h_{125}$.
We know that $h_{125}$
must have some CP-even content because it couples at tree level to $ZZ$.
However, in a theory with CP-violation in the scalar sector (such as the C2HDM), $h_{125}$ could have a mixed CP nature.
This possibility can be probed in the couplings to fermions in a variety of ways.
The simplest case occurs if $c^e_f c^o_f \neq 0$, meaning that in the coupling to some fermion there are both CP-even and CP-odd components, thus establishing CP-violation.
A more interesting case occurs if $h_{125}$ (or some other mass eigenstate) has a pure scalar component to a given type of fermion ($f$)
and a pure pseudoscalar component to a different type of fermion ($g$).
This would render $c^e_f c^o_f = 0 = c^e_g c^o_g$ in a CP-violating model, where $c^e_f c^o_g \neq 0$.
We will now show that in the case of the C2HDM this possibility is no longer available for Type I and it is only available in Type II if $h_{125}= H_2$.
In contrast, the possibility still exists in the Lepton-Specific and Flipped models.
We will also give several benchmark points to allow a more detailed study of these scenarios.

Applying all the above constraints on the parameter space, we have obtained
the points in parameter space that are still allowed. We call this sample 1. We
have also performed a scan with the EDM constraints turned off, which we
call sample 2.
\begin{figure}[t]
  \centering
  \includegraphics[width=0.57\linewidth]{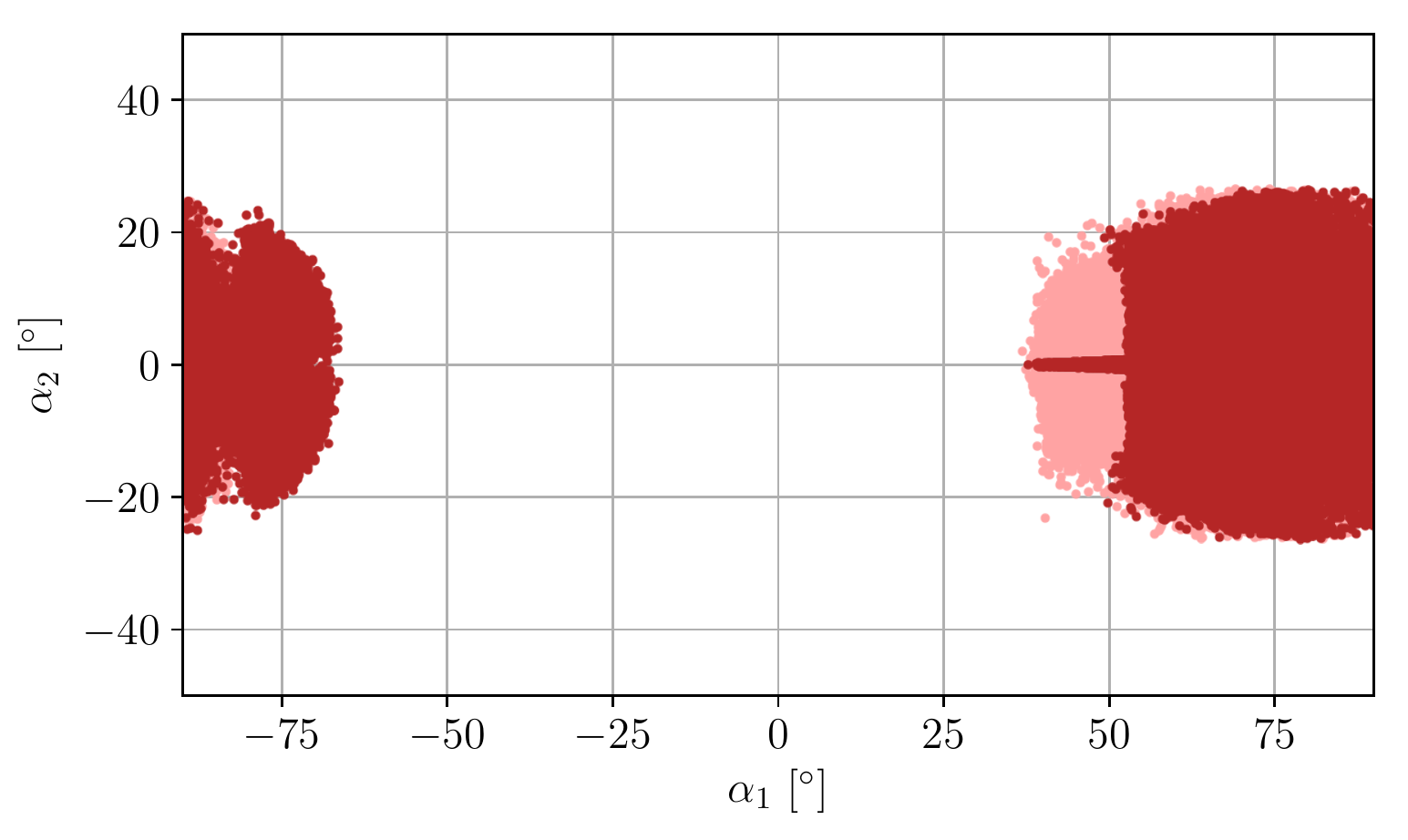}
  \includegraphics[width=0.37\linewidth]{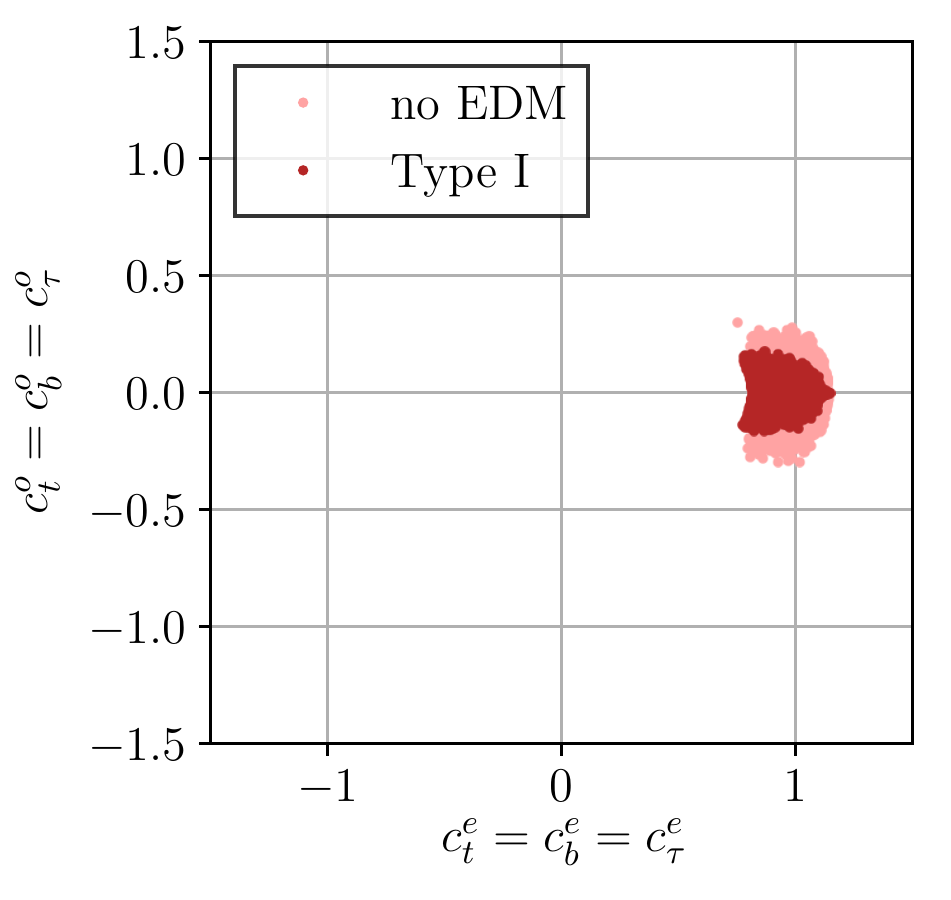}
  \caption{C2HDM Type I: For sample 1 (dark) and sample 2 (light)
    left: mixing angles $\alpha_1$ and $\alpha_2$ of
    the C2HDM mixing matrix $R$ only including scenarios where $H_1=h_{125}$; right: Yukawa
    couplings. } \label{fig:t1c2hdmyuks}
\end{figure}
The left plot of figure~\ref{fig:t1c2hdmyuks} displays the mixing angles $\alpha_1$,
which (in the displayed case of $H_1=h_{125}$) mixes the CP-even parts of the two Higgs doublets and $\alpha_2$,
which parametrises the amount of CP-violating
pseudoscalar admixture, for Type I. In the right plot we present the CP-odd and the CP-even
components of the Yukawa couplings for the Type I model with and without the EDM constraints.
Both plots
clearly show that the EDM constraints have little effect on the
mixing angle $|\alpha_2|$, which can go up to 25$^\circ$ when all constraints
are taken into account.

The maximum value of this angle can be understood from the bound
$0.79 < \mu_{VV} < 1.48$. In fact, as previously shown in \cite{Fontes:2015mea}
the fact alone that $\mu_{VV} > 0.79$ forces the angle $|\alpha _2|$ to be below
$\approx 27^o$. Coming from the bound on $\mu_{VV}$, this constraint will be approximately
the same for all types (before imposing EDM constraints), as will become clear in the next plots.

We are also interested in the wrong-sign regime, defined by a relative sign of the Yukawa
coupling compared to the Higgs-gauge coupling, realized for
$c_b^e < 0$.
As shown previously in
\cite{Ferreira:2014naa, Ferreira:2014dya},
the right plot again demonstrates that the wrong-sign
regime is in conflict with the Type I constraints because
the Yukawa couplings cannot be varied independently.

\begin{figure}[t]
  \centering
  \includegraphics[width=0.55\linewidth]{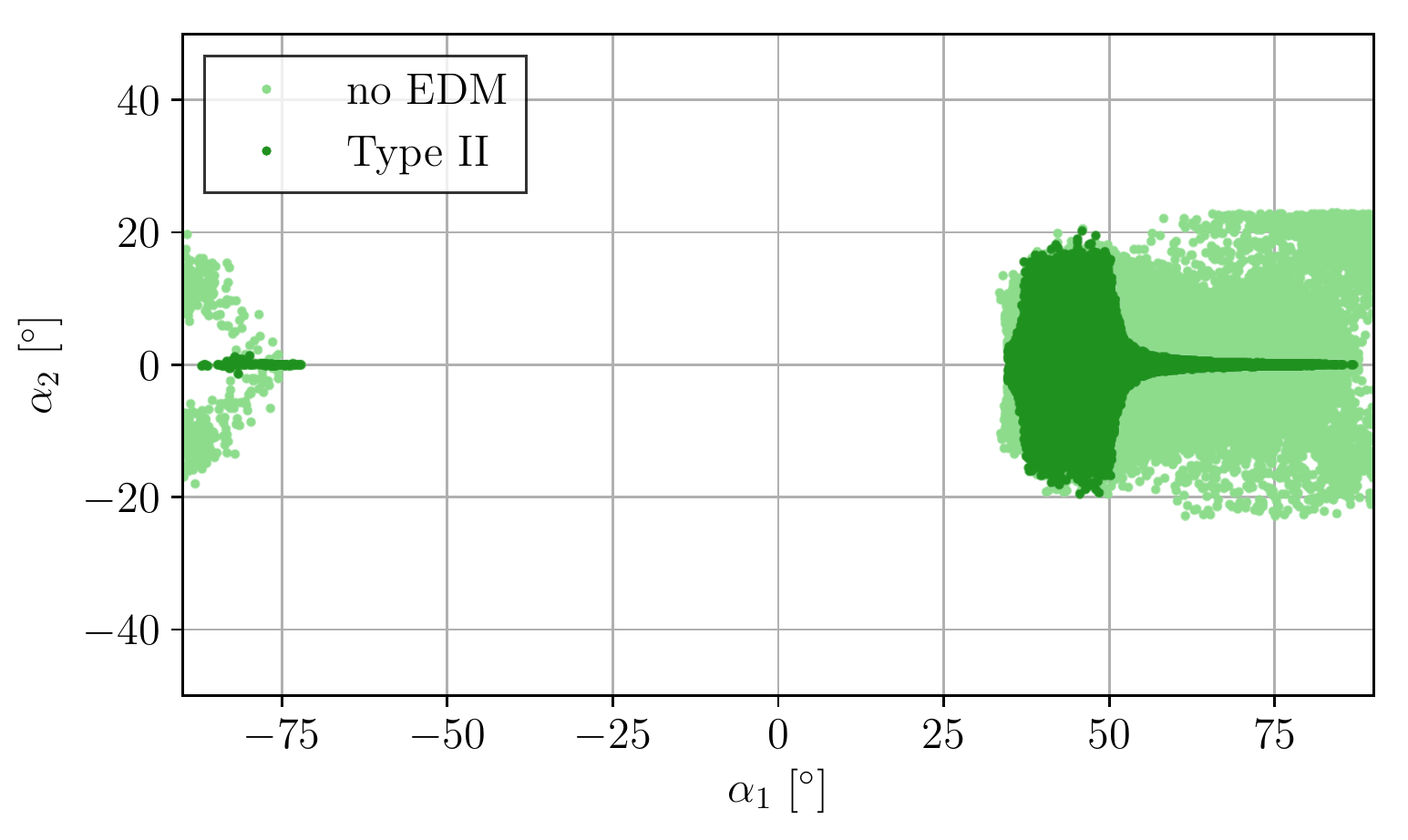}
  \caption{Type II, $H_1=h_{125}$: Mixing angles $\alpha_1$ and $\alpha_2$ of the C2HDM Type
    II mixing matrix $R$ for sample 1 (dark) and sample 2
    (light).}\label{fig:mixangledist}
\end{figure}
In figure~\ref{fig:mixangledist} we present the distributions
of the angle $\alpha_1$ and $\alpha_2$ for samples 1 and 2 and for a Type II model.
The EDM constraints, applied in our sample 1, strongly reduce
$|\alpha_2|$ to small values. Only for scenarios around the maximal
doublet mixing case with $\alpha_1 \approx \pi/4$, $\alpha_2$ can
reach values of up to $\sim \pm 20^\circ$.

\begin{figure}[t]
  \centering
  \includegraphics[width=0.9\linewidth]{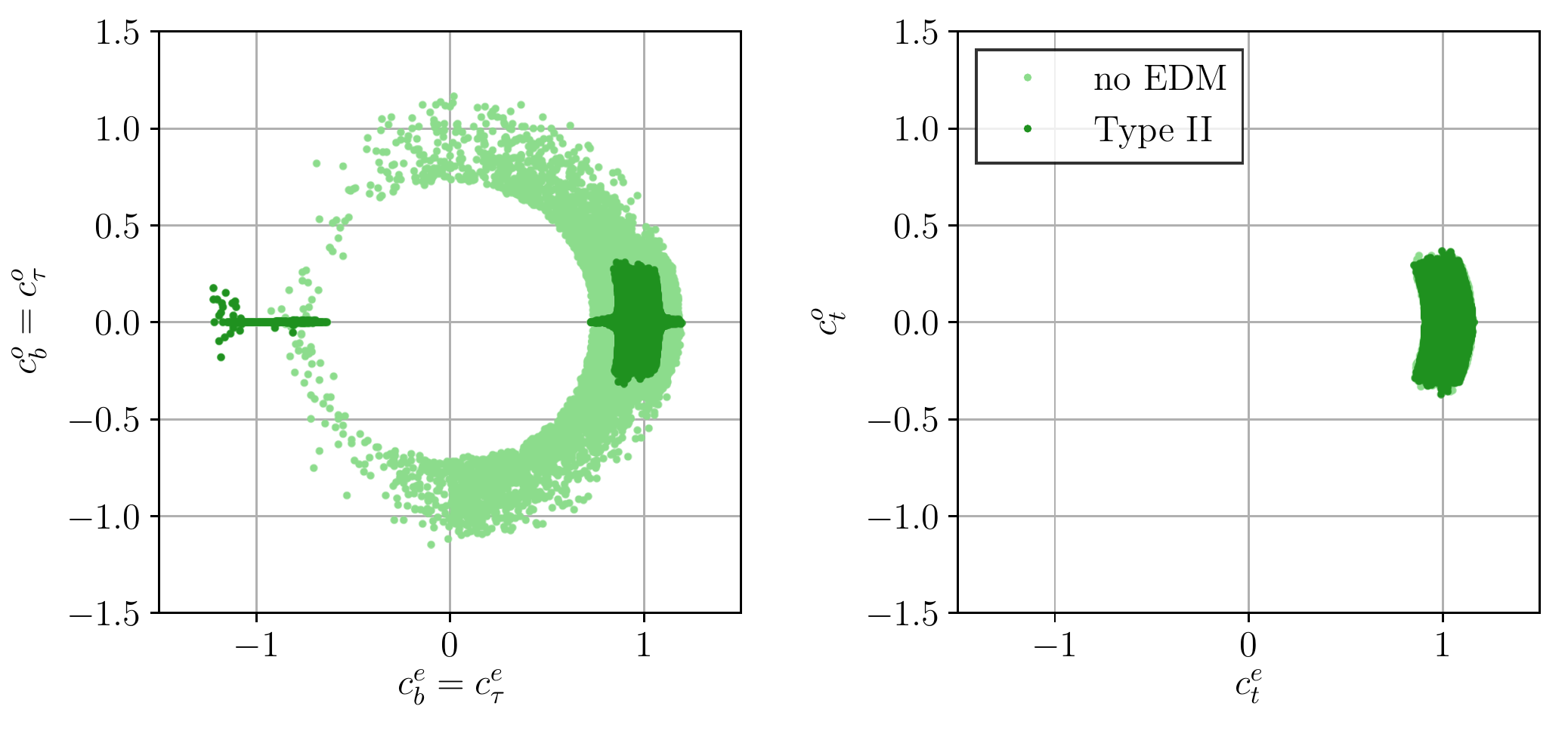}
  \caption{C2HDM Type II, $h_{125}=H_1$: Yukawa couplings to bottom quarks and tau leptons (left) and
    top quarks (right) for sample 1 (dark) and sample 2
    (light). \label{fig:t2c2hdmyuks_H1}}
\end{figure}
The phenomenological implications of the reduced CP-violating mixing
angle in Type II when $h_{125}=H_1$ are demonstrated in
figure~\ref{fig:t2c2hdmyuks_H1}.
It shows the distribution of the CP-odd component $c^o_f$ versus
the CP-even component $c^e_f$ of the $h_{125}$ Yukawa coupling
as defined in eq.~(\ref{eq:yuklag}) to bottom quarks and tau leptons (left)
and top quarks (right).
As can be inferred from figure~\ref{fig:t2c2hdmyuks_H1} (left)
the Higgs data alone still allow for vanishing scalar
couplings to down-type quarks ($c^e_b=0$), as discussed
in \cite{Fontes:2015mea}.
The inclusion of the EDM constraints, however, clearly
rules out this possibility when $h_{125}=H_1$.
Nevertheless,
the wrong-sign regime ($c_b^e < 0$) is still possible in
the C2HDM for down-type Yukawa couplings.
The electron EDM has no discernable effect on the allowed coupling to up-type quarks, as
can be read off from the right plot.

\begin{figure}[t]
  \centering
  \includegraphics[width=0.9\linewidth]{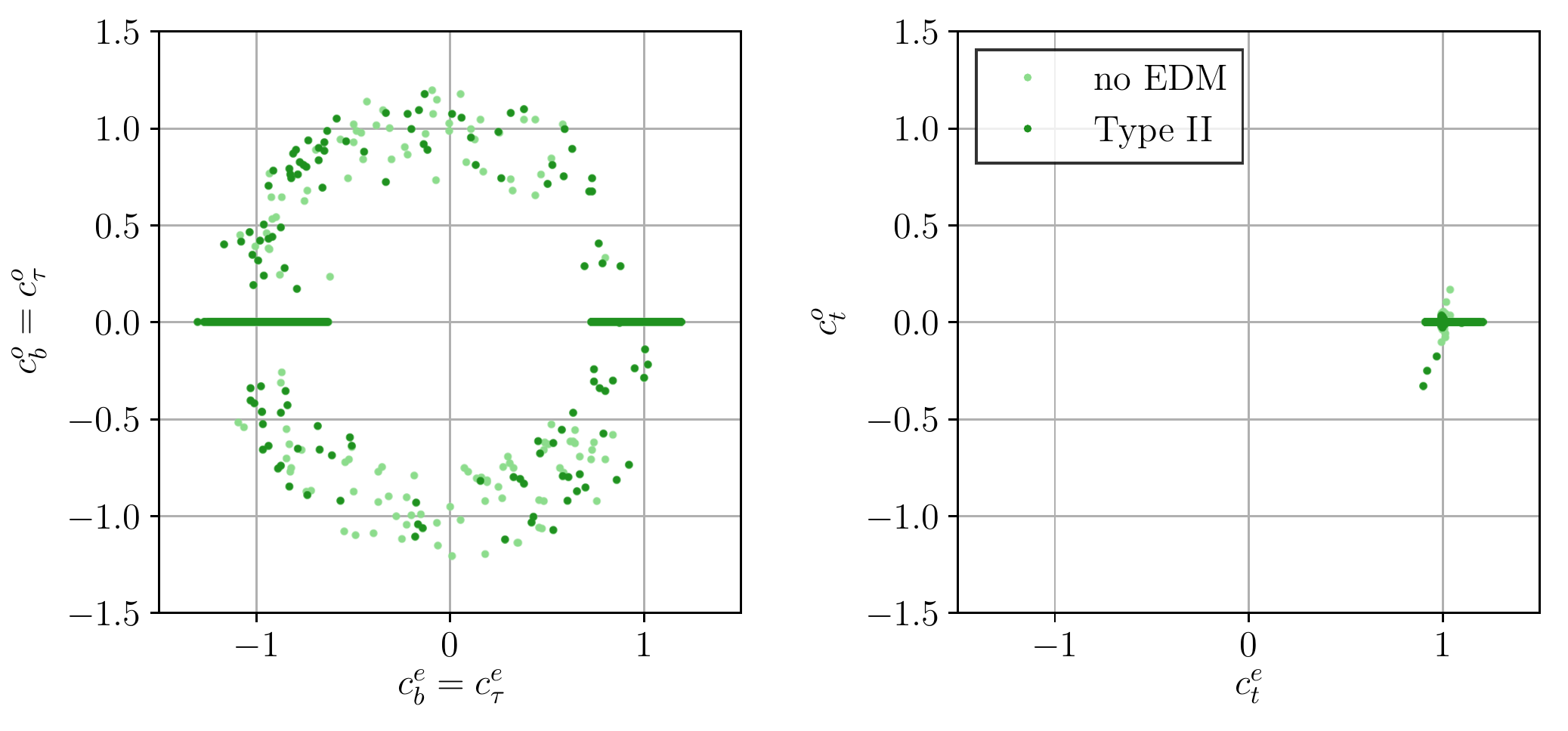}
  \caption{C2HDM Type II, $h_{125}=H_2$: Yukawa couplings to bottom quarks (left) and
    top quarks (right) for sample 1 (dark) and sample 2
    (light). \label{fig:t2c2hdmyuks_H2}}
\end{figure}

The situation changes when we take Type II with $h_{125}=H_2$,
as shown in figure~\ref{fig:t2c2hdmyuks_H2}.
One can still find scenarios
where the top coupling is mostly CP-even ($c_t^e \simeq 1$),
while the bottom coupling is mostly CP-odd ($c_b^o \simeq 1$).
It is noteworthy that the electron EDM kills all such points in Type II
when $h_{125}=H_1$,
but that they are still allowed in Type II when
$h_{125}=H_2$.

In table~\ref{tab:benchII} we present three benchmark scenarios in
Type II with large CP-violation in the Yukawa sector. The first
scenario, BP2m, has maximal $c^o_b$ with nearly vanishing
$c^e_b$. Since $c^e_t$ is always $\simeq1$ this means that $c^e_t c^o_b$
is maximal here. The other two scenarios BP2c and BP2w both have
maximal $c^e_b c^o_b$ but are in the correct sign and wrong-sign
regime, respectively. As discussed above all of the scenarios with
large CP-violation in the Yukawa couplings of $h_{125}$ require
$H_2=h_{125}$ in Type II models.
\begin{table}
  \centering
  \begin{tabular}{lccc}
    \toprule
    Type II & BP2m & BP2c & BP2w\\
    \midrule
    $m_{H_1}$ & 94.187 & 83.37 & 84.883 \\
    $m_{H_2}$ & 125.09 & 125.09 & 125.09 \\
    $m_{H^\pm}$ & 586.27 & 591.56 & 612.87 \\
    $\mathrm{Re}(m_{12}^2)$ & 24017 & 7658 & 46784 \\
    $\alpha_1$ & -0.1468 & -0.14658 & -0.089676 \\
    $\alpha_2$ & -0.75242 & -0.35712 & -1.0694 \\
    $\alpha_3$ & -0.2022 & -0.10965 & -0.21042 \\
    $\tan\beta$ & 7.1503 & 6.5517 & 6.88 \\
    \midrule
    $m_{H_3}$ & 592.81 & 604.05 & 649.7 \\
    $c^e_b=c^e_\tau$ & 0.0543 & 0.7113 & -0.6594 \\
    $c^o_b=c^o_\tau$ & 1.0483 & 0.6717 & 0.6907 \\
    \midrule
    $\mu_V/\mu_F$ & 0.899 & 0.959 & 0.837 \\
    $\mu_{VV}$ & 0.976 & 1.056 & 1.122 \\
    $\mu_{\gamma\gamma}$ & 0.852 & 0.935 & 0.959 \\
    $\mu_{\tau\tau}$ & 1.108 & 1.013 & 1.084 \\
    $\mu_{bb}$ & 1.101 & 1.012 & 1.069 \\
    \bottomrule
  \end{tabular}
  \caption{Benchmark points with large pseudoscalar Yukawa couplings
    in Type II, $h_{125}=H_2$. Lines 1-8 contain the input parameters; lines 9-11 the
    derived third Higgs boson mass and the relevant Yukawa couplings (multiplied by $\mathrm{sgn}(c(h_{125}VV))$) and the last five lines the signal strengths of $h_{125}$.}
  \label{tab:benchII}
\end{table}

%

\begin{figure}[t]
  \centering
  \includegraphics[width=0.86\linewidth]{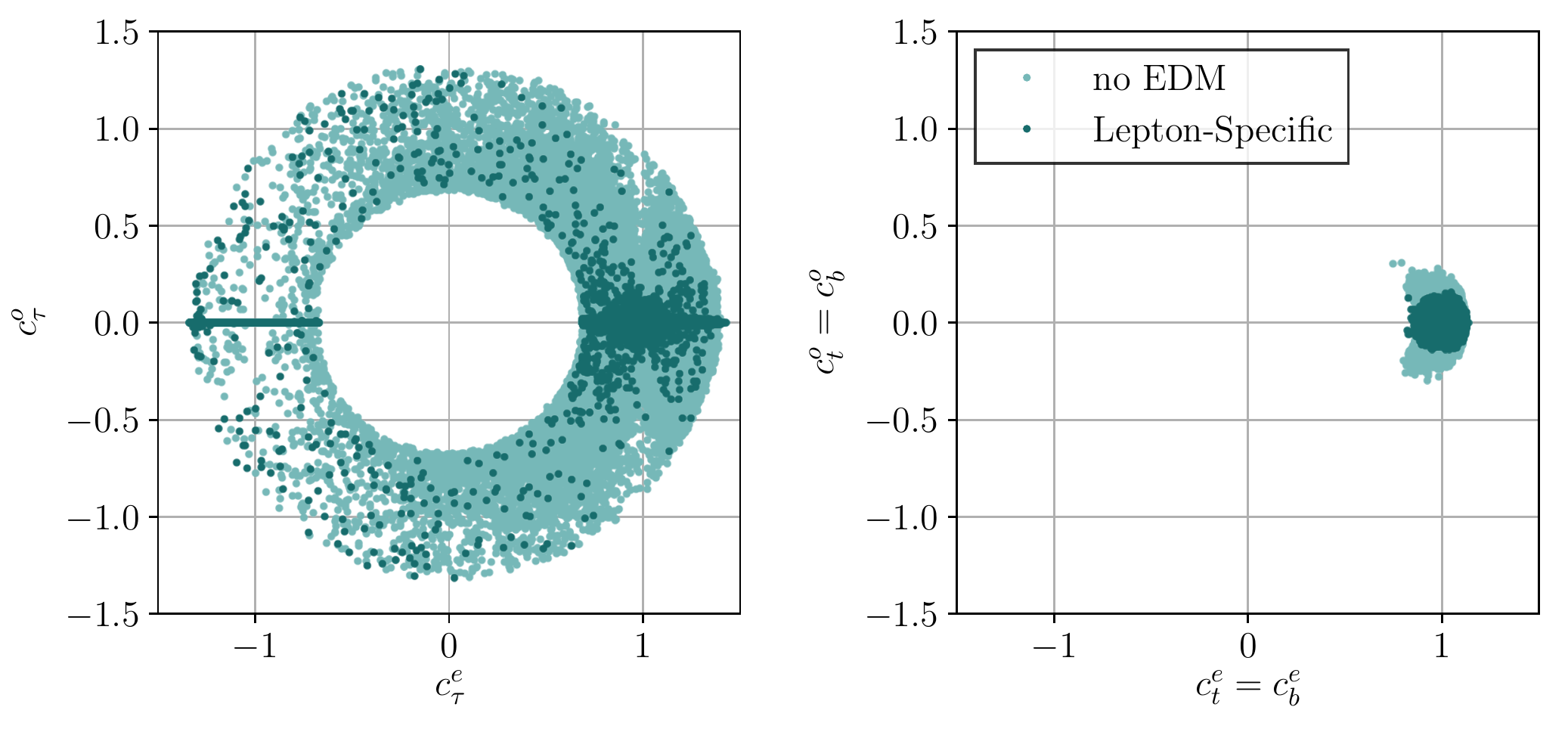}
  \caption{C2HDM Lepton-Specific: Yukawa couplings to charged leptons (left) and
    bottom and top quarks (right) for sample 1 (dark) and sample 2
    (light). \label{fig:lsc2hdmyuks}}
\end{figure}
The situation is even more interesting in the other two Yukawa types.
Figure~\ref{fig:lsc2hdmyuks} displays the Yukawa
couplings for the Lepton-Specific model with and without the EDM constraints.
The down-type quark couplings are tied to the up-type couplings and,
thus, heavily constrained to lie close to the SM (fully CP-even) solution.
However,
figure~\ref{fig:lsc2hdmyuks} (left)
shows that the charged lepton couplings can still be mostly (or even fully)
CP-odd, despite the current EDM constraints.
\begin{figure}[t]
  \centering
  \includegraphics[width=0.86\linewidth]{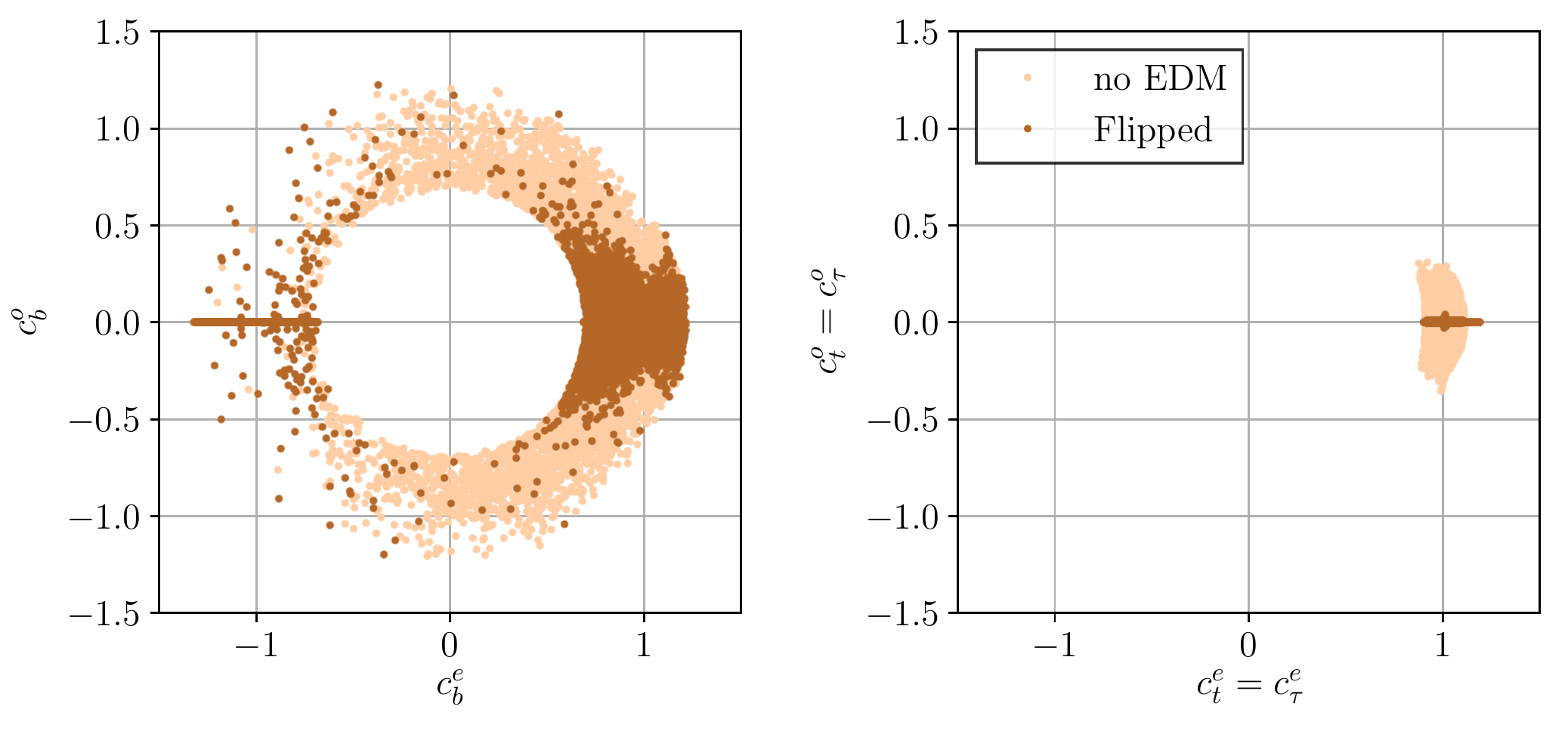}
  \caption{C2HDM Flipped: Yukawa couplings to bottom quarks (left) and
   charged leptons and top quarks (right) for sample 1 (dark) and sample 2
    (light). \label{fig:fc2hdmyuks}}
\end{figure}
Similarly, in the Flipped model the bottom quark can couple to $h_{125}$ in a fully CP-odd fashion, as shown in the left plot
of figure~\ref{fig:fc2hdmyuks}.
In it,
we display the Yukawa
couplings for the Flipped model with and without the EDM constraints.

\begin{figure}[t]
  \centering
  \includegraphics[width=0.55\linewidth]{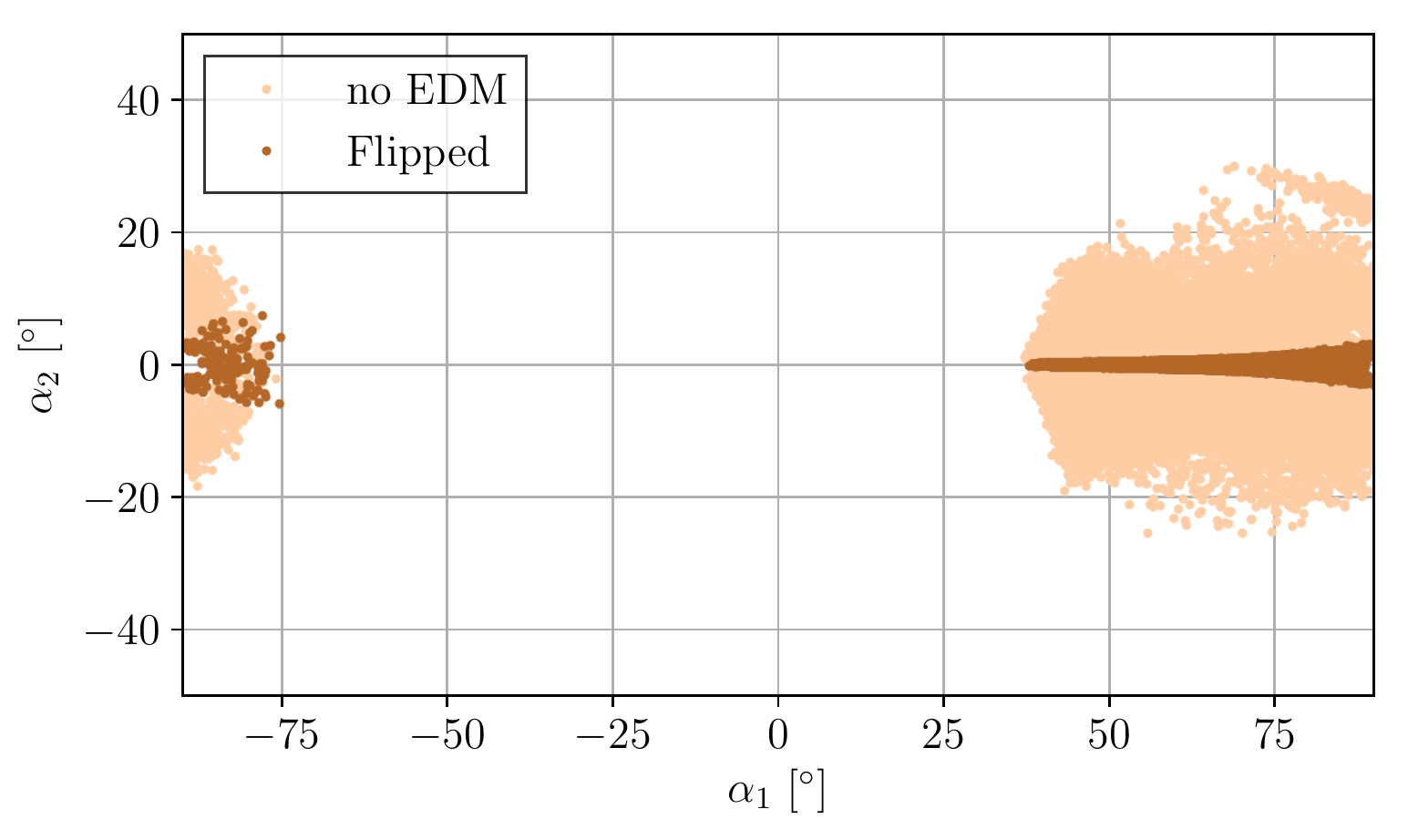}
  \caption{C2HDM Flipped: Mixing angles $\alpha_1$ and $\alpha_2$ of the
	mixing matrix $R$ for sample 1 (dark) and sample 2
    (light).}\label{fig:mixangled_T4}
\end{figure}

As will be discussed below, such large CP-odd components are still viable
in both Lepton-Specific and Flipped models due to cancellations
between the various diagrams entering the EDM calculation. This is
also true for Type II when $H_2 = h_{125}$.
But it is important to stress that they are not due to large $\alpha_2$
values, as illustrated in figure~\ref{fig:mixangled_T4}.
The values for $\alpha_2$ are small,
but $c^o(h_{125} b \bar{b})$ grows very fast as $\alpha_2$
departs from $\alpha_2=0$ for large $\tan{\beta}$.
It grows roughly as $c^o(h_{125} b \bar{b}) \sim s_2 \tan{\beta}$.

\begin{table}[t]
  \centering
\begin{tabular}{lcccclccc}
    \cmidrule[\heavyrulewidth]{1-4}\cmidrule[\heavyrulewidth]{6-9}
    LS & BPLSm & BPLSc & BPLSw & \hspace{2cm}& Flipped & BPFm & BPFc & BPFw\\
    \cmidrule{1-4}\cmidrule{6-9}
    $m_{H_1}$ & 125.09 & 125.09 & 91.619 &&$m_{H_1}$ & 125.09 & 125.09 & 125.09 \\
    $m_{H_2}$ & 138.72 & 162.89 & 125.09 &&$m_{H_2}$ & 154.36 & 236.35 & 148.75 \\
    $m_{H^\pm}$ & 180.37 & 163.40 & 199.29 &&$m_{H^\pm}$ & 602.76 & 589.29 & 585.35 \\
    $\mathrm{Re}(m_{12}^2)$ & 2638 & 2311 & 1651 &&$\mathrm{Re}(m_{12}^2)$ & 10277 & 8153 & 42083 \\
    $\alpha_1$ & -1.5665 & 1.5352 & 0.0110 && $\alpha_1$ & -1.5708 & 1.5277 & -1.4772 \\
    $\alpha_2$ & 0.0652 & -0.0380 & 0.7467 && $\alpha_2$ & -0.0495 & -0.0498 & 0.0842 \\
    $\alpha_3$ & -1.3476 & 1.2597 & 0.0893 && $\alpha_3$ & 0.7753 & 0.4790 & -1.3981 \\
    $\tan\beta$ & 15.275 & 17.836 & 9.870 && $\tan\beta$ & 18.935 & 14.535 & 8.475 \\
    \cmidrule{1-4}\cmidrule{6-9}
    $m_{H_3}$ & 206.49 & 210.64 & 177.52 && $m_{H_3}$ & 611.27 & 595.89 & 609.82 \\
    $c^e_\tau$ & -0.0661 & 0.6346 & -0.7093 && $c^e_b$ & -0.0003 & 0.6269 & -0.7946 \\
    $c^o_\tau$ & 0.9946 & 0.6780 & -0.6460 && $c^o_b$ & -0.9369 & 0.7239 & 0.7130 \\
    \cmidrule{1-4}\cmidrule{6-9}
    $\mu_V/\mu_F$ & 0.980 & 0.986 & 0.954  && $\mu_V/\mu_F$ & 0.927 & 0.964 & 0.844 \\
    $\mu_{VV}$ & 1.014 & 1.029 & 1.000 && $\mu_{VV}$ & 1.154 & 1.091 & 0.998 \\
    $\mu_{\gamma\gamma}$ & 0.945 & 1.018 & 0.879 && $\mu_{\gamma\gamma}$ & 1.027 & 0.986 & 0.874 \\
    $\mu_{\tau\tau}$ & 1.007 & 0.880 & 0.943 && $\mu_{\tau\tau}$ & 1.148 & 1.084 & 1.039 \\
    $\mu_{bb}$ & 1.013 & 1.020 & 1.025 && $\mu_{bb}$ & 1.001 & 0.992 & 1.170 \\
    \cmidrule[\heavyrulewidth]{1-4}\cmidrule[\heavyrulewidth]{6-9}
  \end{tabular}
  \caption{Benchmark points with large pseudoscalar Yukawa couplings in the Lepton-Specific (LS) and Flipped types. Lines 1-8 contain the input parameters; lines 9-11 the derived third Higgs mass and the relevant Yukawa couplings (multiplied by $\mathrm{sgn}(c(h_{125}VV))$) and the last five lines the signal strengths of $h_{125}$.}
  \label{tab:benchLSF}
\end{table}

In table~\ref{tab:benchLSF} we present further benchmark scenarios in
the Lepton-Specific and Flipped types. The BPLSm scenario has a
maximal $c^o_\tau$ coupling with tiny $c^e_\tau$, thus here $h_{125}$
appears CP-even in its couplings to quarks and CP-odd in its couplings to leptons.
In BPLSc and BPLSw the product
$c^e_\tau c^o_\tau$ is maximal.
In the Flipped model we provide three benchmark points. The first one,
BPFm, again having maximal $c^o_b$
while BPFc and BPFw both have large
$c^o_b c^e_b$ but with opposite signs and are therefore close to the
correct and wrong-sign limit, respectively. Note that, in contrast to
the Type II, all benchmark points except BPLSw have $H_1=h_{125}$. We
chose BPLSw with $H_2=h_{125}$ since maximising $c^e_\tau c^o_\tau$ close
to the wrong-sign limit of the Lepton-Specific model
leads to this mass ordering.

\subsection{\label{subsec:H3}The case $H_3=h_{125}$}

One interesting possibility in the C2HDM is to have the 125 GeV scalar
discovered at LHC ($h_{125}$) coincide with the heaviest Higgs boson ($H_3$).
This possibility is excluded for Type II and Flipped, as
the $B$-physics constraints impose that the charged Higgs boson must be
quite heavy, $m_{H^+}>580$ GeV. This poses a problem with the
electroweak precision tests, specially with the $T$ parameter. Indeed, the way to
accommodate the experimental bounds on $T$ is to have a spectrum that
has some degree of degeneracy. Requiring $m_{H^+}>580$ GeV implies that
the other Higgs boson masses cannot all be below 125 GeV.
Thus, $m_{H^+}>580$ GeV is not compatible with
$m_{H_2}< 125$ GeV.

\begin{figure}[t]
  \centering
  \includegraphics[width=0.9\linewidth]{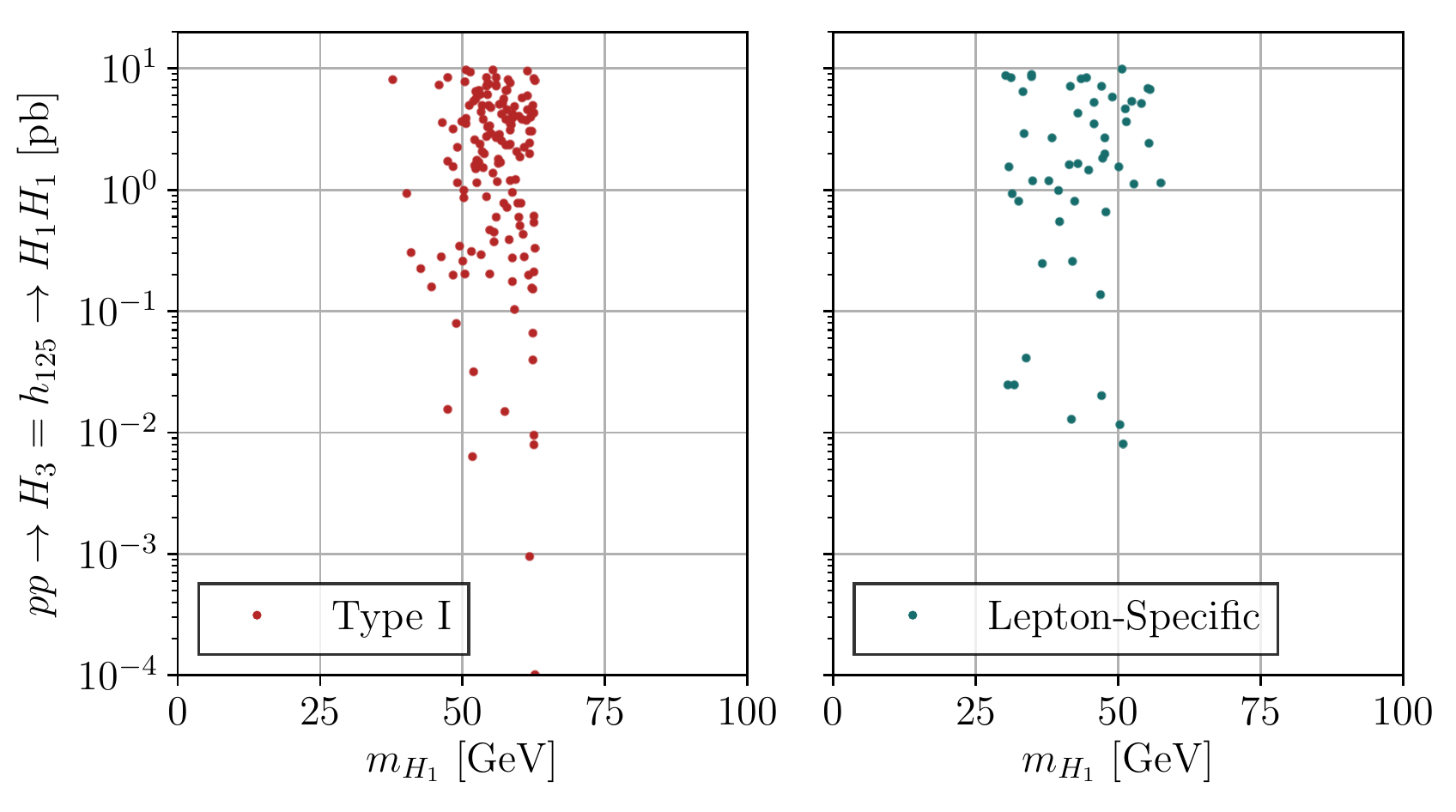}
  \caption{Rate for $pp \to H_3(=h_{125}) \to H_1 H_1$. There are only allowed points for Type I (left)
  and for Lepton-Specific (right).\label{fig:rates_HsmEqH3_HlHl}}
\end{figure}

However, $H_3=h_{125}$ is feasible for Type I and Lepton-Specific,
as in these cases  $B$-physics constraints only impose $m_{H^+}>100$ GeV
(as explained, for low $\tan\beta$ this bound could be slightly higher).
The situation here is quite fascinating,
because it highlights an interesting complementarity
between LHC and the old LEP results.
The relevant Feynman rules are:
\begin{eqnarray}
&&
[H_k Z_\mu Z_\nu]:\ \
i g_{\mu \nu} \frac{g}{c_W} m_Z\, c(H_k Z Z),
\nonumber\\
&&
[H_i H_j Z_\mu]:\ \
i \frac{g}{2 c_W} (p_i - p_j)_\mu\, c(H_i H_j Z),
\end{eqnarray}
where $p_i$ and $p_j$ are the incoming momenta of particles $H_i$ and $H_j$, respectively,
$c(H_k Z Z)$ is given in eq.~(\ref{eq:c2dhmgaugecoup}), and
\begin{equation}
c(H_i H_j Z) = \epsilon_{ijk}\,  c(H_k Z Z).
\end{equation}
In the SM,
$c(H_k Z Z)=1$ and $c(H_i H_j Z) = 0$.
Equating  $H_3=h_{125}$,
the LHC $h_{125} \to ZZ$ signal forces $c(H_3 Z Z) \sim 1$.
But this means that $c(H_1 H_2 Z) \sim 1$ and,
if $m_{H_3} > m_{H_1} + m_{H_2}$,
then the decay $Z \to H_1 H_2$ would have been seen at LEP \cite{Schael:2006cr}.
Thus, all points with $H_3=h_{125}$ must have
$m_{H_3} < m_{H_1} + m_{H_2} \leq 2  m_{H_2}$. If $m_{H_3} > 2 m_{H_1}$
the decay $H_3 \to H_1 H_1$ is possible.
This decay is still possible in Type I and Lepton-Specific,
as shown in figure~\ref{fig:rates_HsmEqH3_HlHl}. In fact, the rates can go
up to about 10 pb and therefore have excellent prospects of being probed
at the LHC Run2. If we include the decays of the $H_1$ we retain cross sections of up to 1 pb in the $H_3 \to H_1 H_1 \to b \bar{b} \tau^+ \tau^-$
channel in both models and up to 50 fb in the $H_3 \to H_1 H_1 \to b \bar{b} \gamma\gamma$ channel in Type I.
Notably, even within this subset of points, one can still find Lepton-Specific
corners of parameter space which obey EDM constraints and still allow for
large CP-odd components in $c^o(h_{125} \tau \bar{\tau})$.

\section{Measures of CP-violation}
 \label{sec:measure}

Throughout this section we will use the notation $H_\downarrow$ ($H_\uparrow$)
to designate the lightest (heaviest) of the non-$h_{125}$ neutral Higgs bosons.
Their mass can be below or above 125 GeV.
The manifestation of CP-violation in models with two Higgs doublets can be probed
with a number of variables even if there is only one independent CP-violating
phase in the scalar sector. The most obvious variable is the phase in one of the
complex parameters of the potential, that is, either $\lambda_5$  or $m_{12}^2$.
As the two phases are not independent,
we have opted to use $\phi (\lambda_5)$.
As discussed in \cite{Branco:1999fs, Fontes:2015xva},
there are several combinations of Higgs decays that are a clear signal
of CP-violation in any extension of the SM.
Others, like the simultaneous observation of the three decays
$H_i \to ZZ$, $i=1,2,3$,
enable us to distinguish the 2HDM from
the C2HDM but are not unequivocal signs of CP-violation in general extensions of the SM.
The question we are trying to address now is: are there variables that quantify CP-violation from a theoretical point of view
and also correlate with the rates of a combination of decays which would establish CP-violation experimentally?
\begin{figure}[t]
  \centering
  \includegraphics[width=0.62\linewidth]{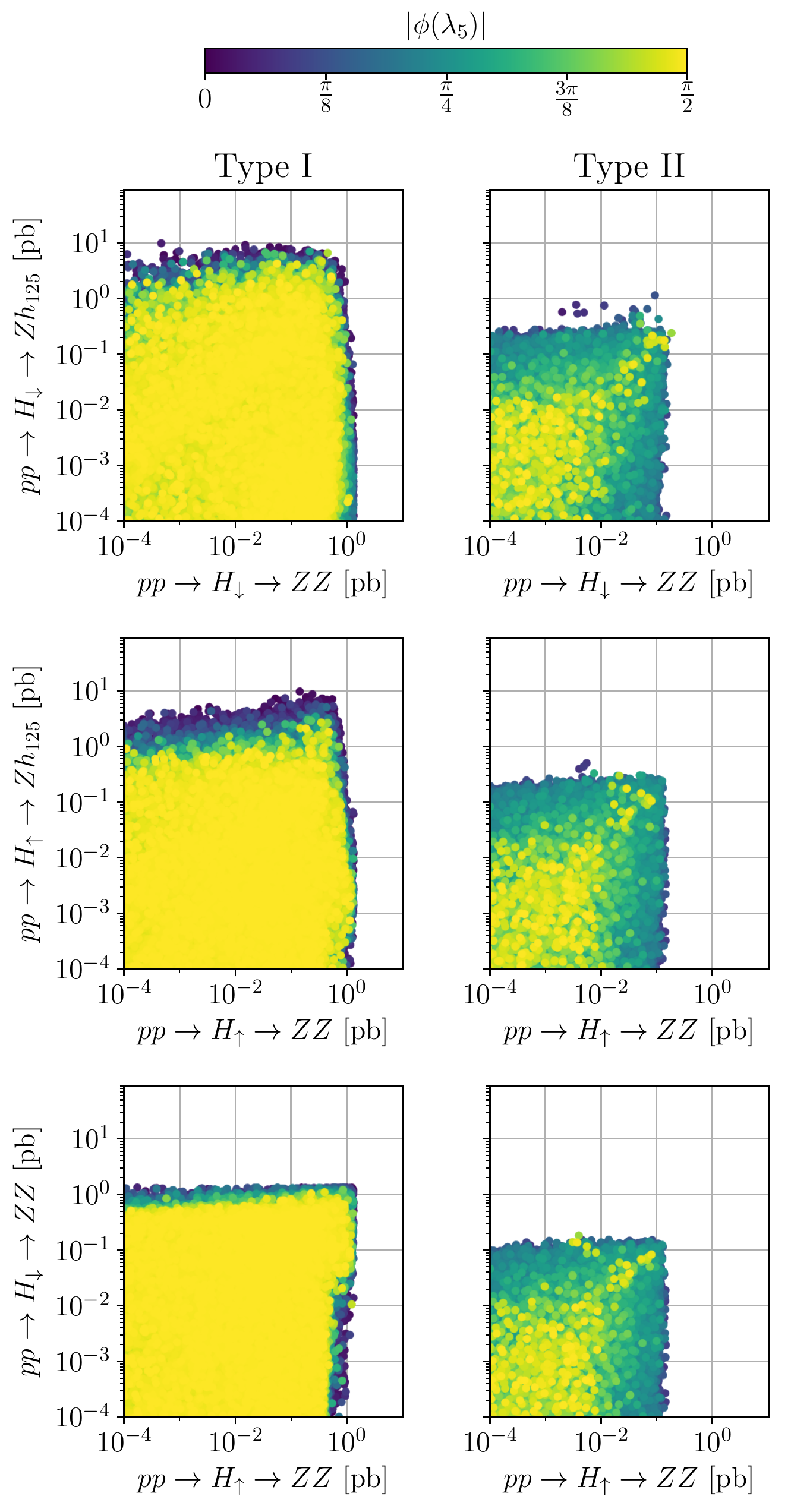}
  \caption{Set of CP-violating processes as a function of the CP-violation phase $|\phi (\lambda_5)|$
  (see colour code) for
  Type I (left column) and Type II (right column).
  In the first row we show $pp \to H_\downarrow \to Z h_{125}$ against $pp \to H_\downarrow \to ZZ$, in the second row
  we have $pp \to H_\uparrow \to Z h_{125}$ against $pp \to H_\uparrow \to ZZ$ and in the third row we plot
  $pp \to H_\downarrow \to ZZ$ against $pp \to H_\uparrow \to ZZ$. Note that the yellow points are superimposed
  on the darker points - there are dark points underneath the yellow points.} \label{fig:CPvarangle}
\end{figure}

Starting with
$\phi (\lambda_5)$, we present in figure~\ref{fig:CPvarangle} three classes of CP-violating processes
as a function of the CP-violation phase $|\phi (\lambda_5)|$ for
  Type I (left column) and Type II (right column).
  In the first row we show $pp \to H_\downarrow \to Z h_{125}$  against $pp \to H_\downarrow \to ZZ$, in the second row
  we have $pp \to H_\uparrow \to Z h_{125}$ against $pp \to H_\uparrow \to ZZ$ and in the third row we plot
  $pp \to H_\downarrow \to ZZ$ against $pp \to H_\uparrow \to
  ZZ$. These classes of decays were chosen because
  together with the already observed process $h_{125} \to ZZ$ they can be used to identify CP violation, and also because searches for the other processes
  were performed for Run 1 and will continue during Run 2.
  There are two striking features in the plots. First,
  the production rates in Type I are almost one order of magnitude above the ones for Type II. This is
  because there are constraints that act more strongly on Type II like $b \to s \gamma$ or the EDM constraints
  as we will see later.
  Ultimately, it is due to the different structure of the Yukawa couplings: in Type
  I all Yukawa couplings are equal, making the model harder to constrain.
  Second, there is no correlation between the magnitude
  of $\phi (\lambda_5)$ and the production rates, because the points with larger
  $\phi (\lambda_5)$ are
  almost evenly spread throughout the plot. It is not that we expected that large values of the
  CP-violating phase would correspond to large production rates for any of the processes in the plot.
  In fact, maximal CP-violation is attained for specific sets of values of the angles but the production
  rates are a complicated combination of all parameters of the model. Still some colour structure could have
  emerged in the plots, but this was not the case.

In view of the negative result for the complex phase, we have looked for other variables \cite{Mendez:1991gp, Khater:2003ym}
that could be used as a measure of CPV in the production and decay of the three neutral Higgs bosons
\cite{Branco:1999fs, Fontes:2015xva}. The sets of variables proposed in the literature are basically of two types: the ones
where the CP-violating variables appear in a sum of squares,
and the ones where they appear in a product.
The important difference between them is that while the former is zero
only when there is no CP-violation in the model, the latter can be
zero even if the model is CP-violating.
However, if CP is conserved, both variables are zero.

We start by defining a multiplicative variable first proposed in \cite{Mendez:1991gp} that
allows us to distinguish a CP-conserving from a CP-violating 2HDM,
\begin{equation} \label{Eq:xi-v}
\xi_V=27[g_{h_1VV}\,g_{h_2VV}\,g_{h_3VV}]^2 = 27\prod_{i=1}^3[\cos\beta R_{i1}+\sin\beta R_{i2}]^2 = 27\prod_{i=1}^3T_{1i}^2 \, ,
\end{equation}
which, when the couplings $g_{VVH_i}$ are normalised to the SM one, satisfies
\begin{equation}
0\le \xi_V \le 1.
\end{equation}

With the purpose of finding quantities invariant under a basis transformation
that change sign under a CP transformation, it was shown in \cite{Lavoura:1994fv}
that the simplest CP-odd invariant that can be built from the mass matrix is
\begin{equation} \label{Eq:J1}
J_1= (m_1^2 - m_2^2) (m_1^2-m_3^2) (m_2^2-m_3^2) \prod_{i=1}^3T_{1i} \, .
\end{equation}
Furthermore, any other CP-odd invariant built from the mass matrix alone has
to be proportional to $J_1$ \cite{Lavoura:1994fv}. There is a relation between
$\xi_V$ and $J_1^2$ that can be written as
\begin{equation} \label{Eq:J1a}
J_1^2= [(m_1^2 - m_2^2) (m_1^2-m_3^2) (m_2^2-m_3^2)]^2 \frac{\xi_V}{27}\, .
\end{equation}
It should be noted that even if $J_1=0$, CP-violation could occur in the
scalar sector (see
\cite{Lavoura:1994fv} for details).

\begin{figure}[t]
  \centering
  \includegraphics[width=0.8\linewidth]{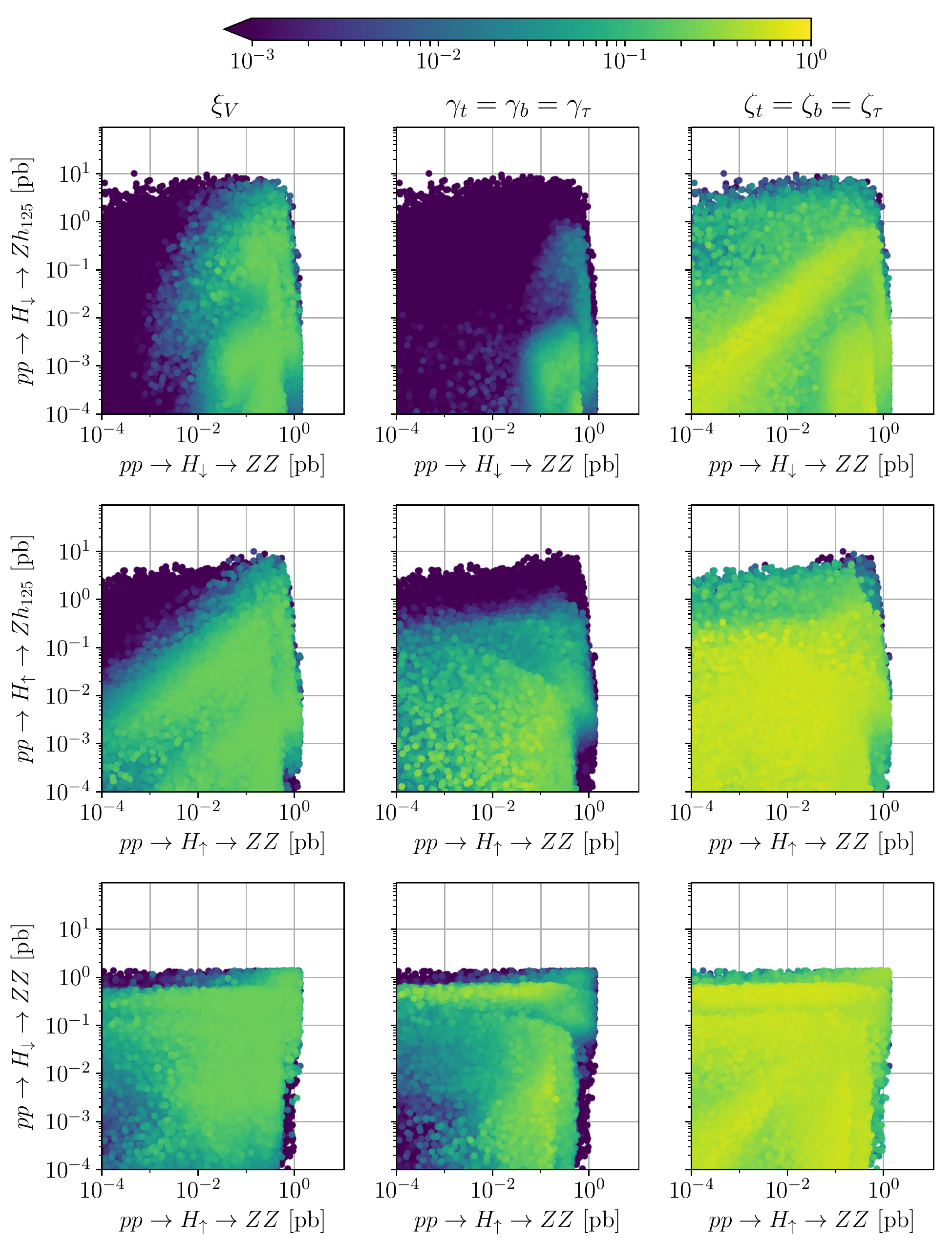}
  \caption{Classes of CP-violating processes as a function of the CP-violating variables (see colour code)
   for the Type I C2HDM.
  In the first row we show $pp \to H_\downarrow \to Z h_{125}$ against $pp \to H_\downarrow \to ZZ$, in the second row
  we have $pp \to H_\uparrow \to Z h_{125}$ against $pp \to H_\uparrow \to ZZ$ and in the third row we plot
  $pp \to H_\downarrow \to ZZ$ against $pp \to H_\uparrow \to ZZ$. In each column we show the variable that is being probed. The darker points are underneath
  the lighter ones.} \label{fig:CPvar1}
\end{figure}

This measure of CP-violation is not applicable to the fermion--Higgs sector,
where the invariants of ref.~\cite{Botella:1994cs} apply instead.
Variables that are a clear signal of CP-violation can be built with
the scalar and pseudoscalar components of the Yukawa couplings.
In fact, if a model has CP-violating scalars at tree level
its Yukawa couplings have the general form $c^e_f + i c^o_f \gamma_5$.
Thus, as discussed at the beginning of subsection~\ref{subsec:constraints},
variables of the type $c^e c^o$ clearly signal CP-violation in the model.
Therefore,
we define the normalised multiplicative variables \cite{Khater:2003ym}
\begin{equation} \label{Eq:gamma_tb}
\gamma_t=1024\prod_{i=1}^3[R_{i2}\,R_{i3}]^2, \quad
\gamma_b=1024\prod_{i=1}^3[R_{i1}\,R_{i3}]^2,
\end{equation}
satisfying
\begin{equation}
0\le\gamma_t\le1, \quad 0\le\gamma_b\le1,
\end{equation}
as measures of CP-violation in the up- and down-quark sectors, respectively.
The corresponding normalised sum variables \cite{Khater:2003ym} are defined as
\begin{equation}
\zeta_t=2\sum_{i=1}^3[R_{i2}\,R_{i3}]^2, \quad \zeta_b=2\sum_{i=1}^3[R_{i1}\,R_{i3}]^2 \, ,
\end{equation}
with
\begin{equation}
\quad 0 \le \zeta_t \le 1 , \, \qquad  0 \le \zeta_b \le 1 ,
\end{equation}
again for the up- and down-quark sectors, respectively.

We will use the same combinations of decays as in figure~\ref{fig:CPvarangle} to see if any of the variables proposed can provide a relation between the amount of CP-violation and the production rates in case these processes are observed at the LHC.
In figure~\ref{fig:CPvar1} we present the three classes that signal CP-violation as a function of the CP-violation variables for the Type I C2HDM. Since $h_{125} \to ZZ$ (not shown) was already observed, any pair of two processes appearing in the plots combined with the latter form a signal of CP-violation. In the first row we show $pp \to H_\downarrow \to Z h_{125}$ against $pp \to H_\downarrow \to ZZ$, in the second row   we have $pp \to H_\uparrow \to Z h_{125}$ against $pp \to H_\uparrow \to ZZ$ and in the third row we plot $pp \to H_\downarrow \to ZZ$ against $pp \to H_\uparrow \to ZZ$. In each column, we show the variable that is being probed. For the case of Type I, the top and bottom Yukawa couplings are the same. The general picture is that there is no striking correlation between the large values for the variables (more yellow points) and the large production cross sections for each process. There is a quite even spread of yellow points for the sum variable, $\zeta_f$.
This is because even if the product of scalar/pseudoscalar components of the SM-like Higgs boson is very constrained, any of the products of the other two Higgs bosons can be very large, yielding a large value for the sum in $\zeta_f$.
Therefore, variables of this type can always be large, and we will not show them in the remaining plots. Regarding the variables $\gamma_f$, we can see some structure in the plots as there are cases where more yellow points are clustered closer to the maximum values of the production rates.
This trend is clearer in the last row where kinematics play a less important role. In fact, the first row is the most constrained regarding the phase space available while the last row is almost symmetric regarding the reduction of the phase space.
The difference between the first and second row regarding kinematics is just that the first row deals with the lightest non-125 Higgs boson while the second row deals with the heaviest one.

\begin{figure}[t]
  \centering
  \includegraphics[width=0.8\linewidth]{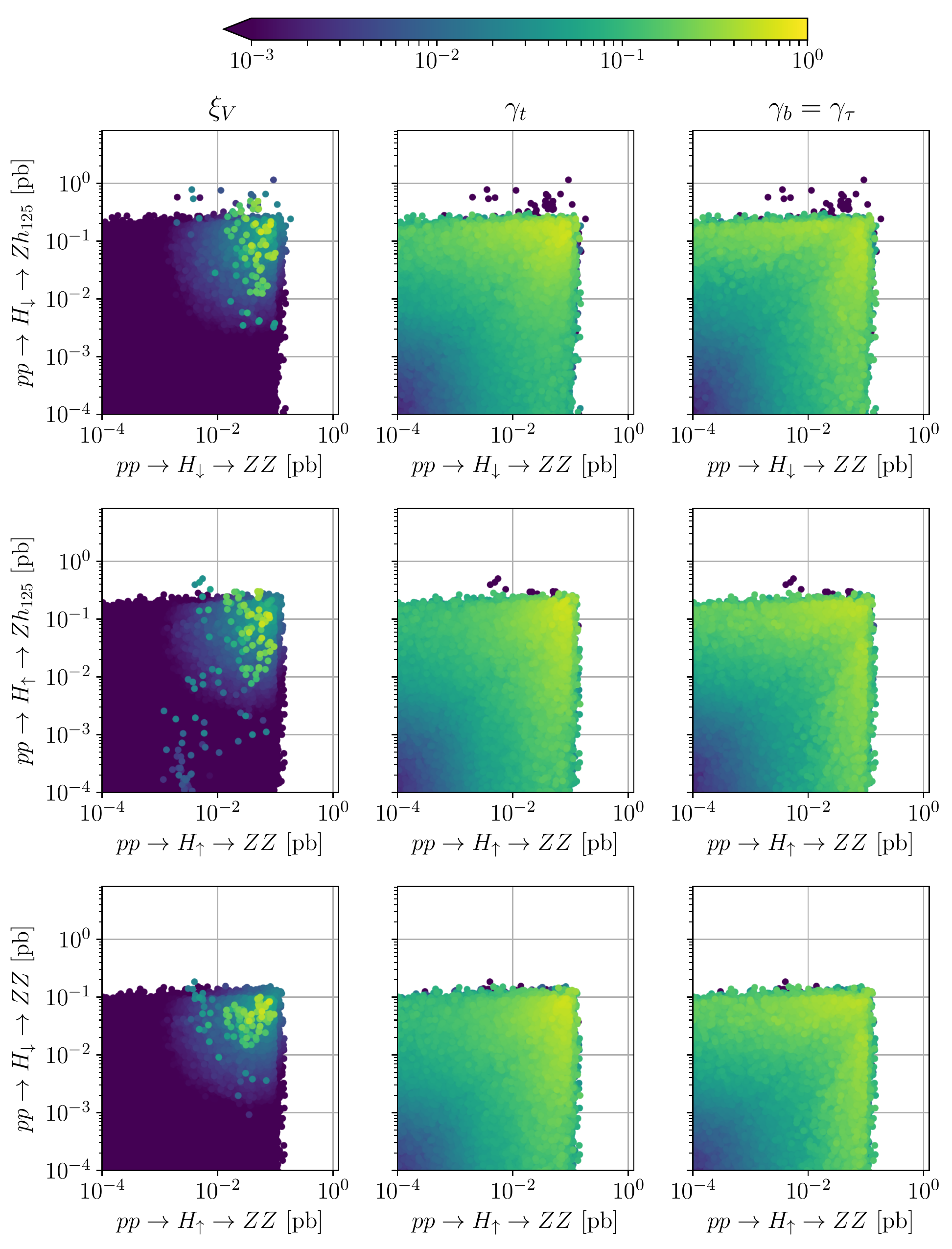}
  \caption{Same as figure~\ref{fig:CPvar1}, but for Type II.} \label{fig:CPvar2}
\end{figure}

In figure~\ref{fig:CPvar2} we present the same three classes of CP-violating processes as a function of CP-violating variables for the Type II C2HDM. What we see for this Yukawa type is that the distribution of the yellow points is more structured and more clustered in specific regions.
In fact, the yellow points tend to cluster more in the regions where the production rates are larger. This behaviour is more striking in the first column, for the variable $\xi_V$, where all yellow points are in the parameter region where the production rates are maximal. For the remaining variables, the distribution of yellow points is again not so structured. Still, all variables are larger when the production rates are both large and are much smaller when both production rates are smaller. However, in all plots, there are always points with small values of the CP-violating variables and large values of the production rates. Hence, although the variables show us a trend, they are not conclusive as a measure of CP-violation in the scalar sector.

\begin{figure}[t]
  \centering
  \includegraphics[width=0.8\linewidth]{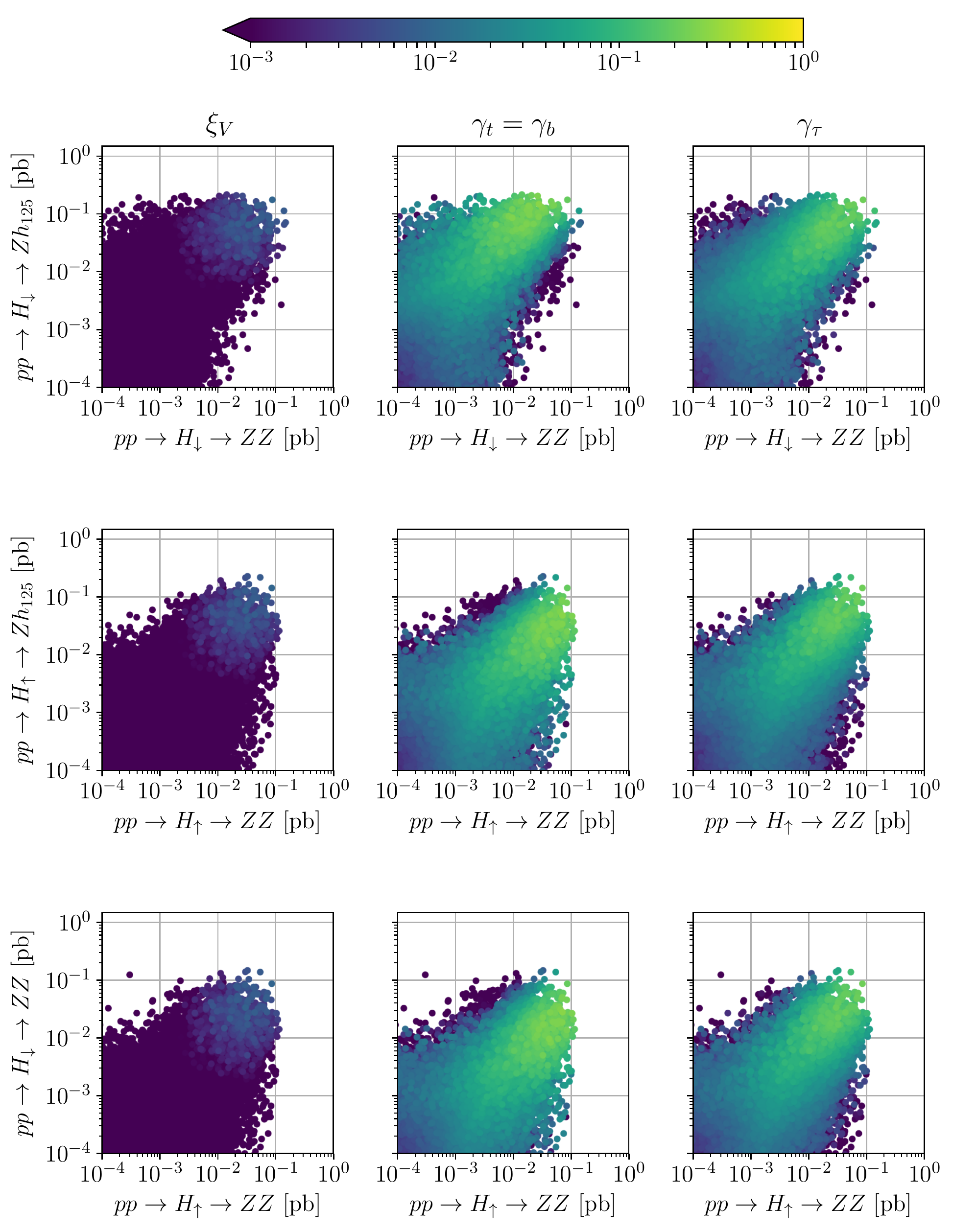}
  \caption{Same as figure~\ref{fig:CPvar2}, but with signal strengths
    within 5\% of the SM values, for Type II.} \label{fig:CPvar3}
\end{figure}
In figure~\ref{fig:CPvar3} we present the same classes in the Type II C2HDM as in figure~\ref{fig:CPvar2}, but with signal strengths (see \eqref{exp_constraints}) within 5\% of the SM values. This gives us a hint on what to expect at the end of the LHC Run2, or at the high luminosity LHC.
There is a clear effect in reducing the production rates but not in the distribution of yellow points.
The main difference is that now no yellow points appear in the first column which means that the points with very large rates were excluded. The distribution of points in the other columns did not change significantly, but the points with the higher rates were also excluded as for the first row. In conclusion, there seems to be an overall reduction in the parameter space of the model leading to smaller production rates.

\begin{figure}[t]
  \centering
  \includegraphics[width=0.8\linewidth]{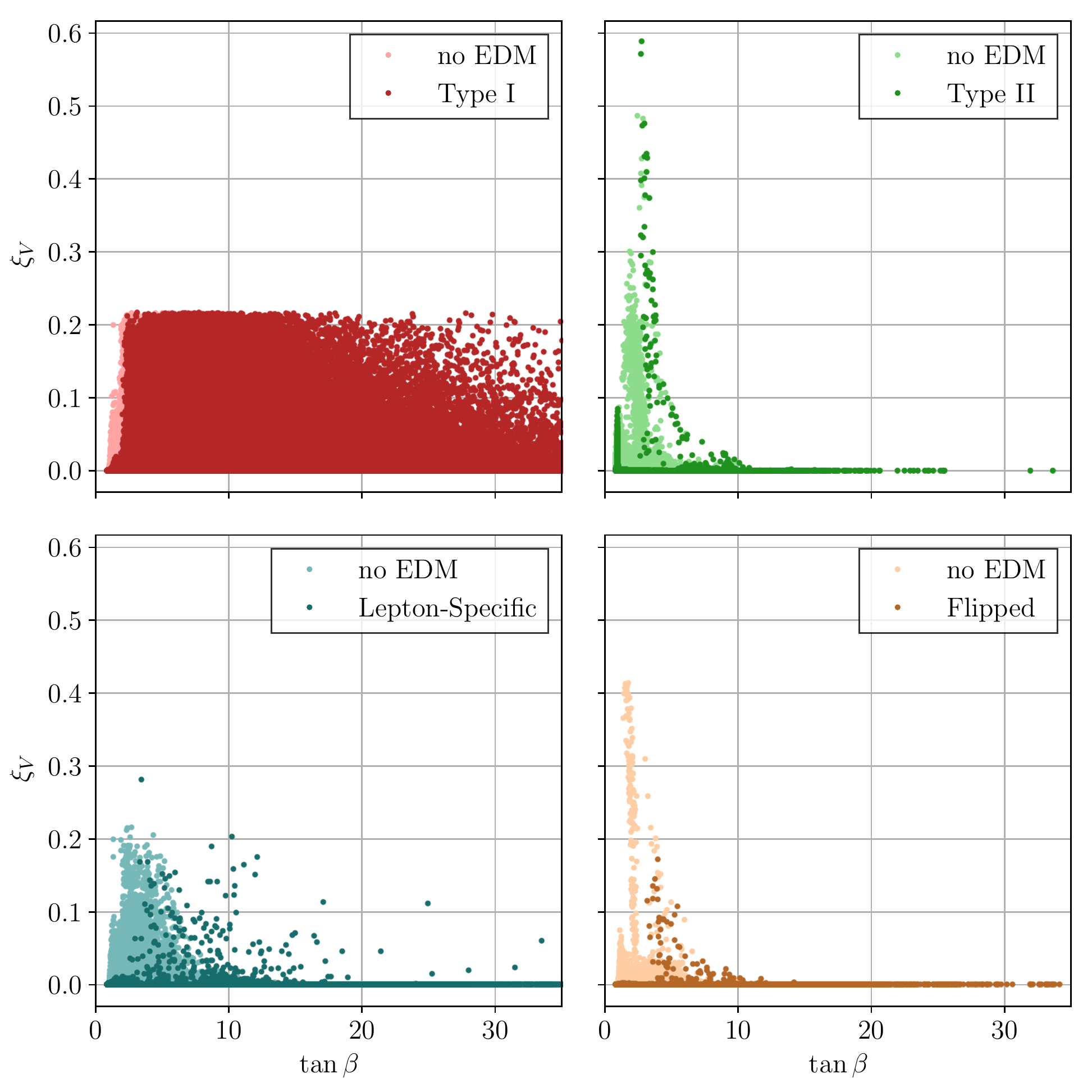}
  \caption{The CP-violating parameter $\xi_V$ as a function of $\tan \beta$ for
  Type I (top left), Type II (top right), Lepton-Specific (bottom left) and
Flipped (bottom right).
  The lighter points have passed all the constraints except for the EDM bounds while the darker
  points have passed all constraints.} \label{fig:xitanbeta}
\end{figure}
In figure~\ref{fig:xitanbeta} we show the CP-violating parameter $\xi_V$ as a function of $\tan \beta$ for Type I (top left), Type II (top right), Lepton-Specific (bottom left) and Flipped (bottom right).
The lighter points have passed all the constraints except for the EDM bounds, while the darker points have passed all constraints. In Type I there is not much difference between the two sets of points, and there are no special regions regarding the allowed values of $\tan \beta$.
Also, the maximum value for $\xi_V$ is around 0.2 almost independently of $\tan \beta$.
For Type II, the results are much more striking. After imposing the EDM constraints, we end with two almost straight lines (one for $\tan \beta \approx 1$ and the other for $\xi_V \approx 0$), as well as a region around $\tan \beta \approx 3$ permitting values of $\xi_V$ up to 0.6. This means that $\tan \beta$ can only be large when we approach the CP-conserving limit except for a few points, which lie in the wrong-sign regime. Hence, in a Type II model, points with significant CP-violation can occur for $\tan \beta \approx 1$ in the alignment limit or for large $\tan \beta$ for the wrong sign limit.
The situation in Flipped is similar to Type II, with a lower maximum value of $\xi_V \sim 0.2$ after imposing the EDM constraints.


\begin{figure}[t]
  \centering
  \includegraphics[width=0.85\linewidth]{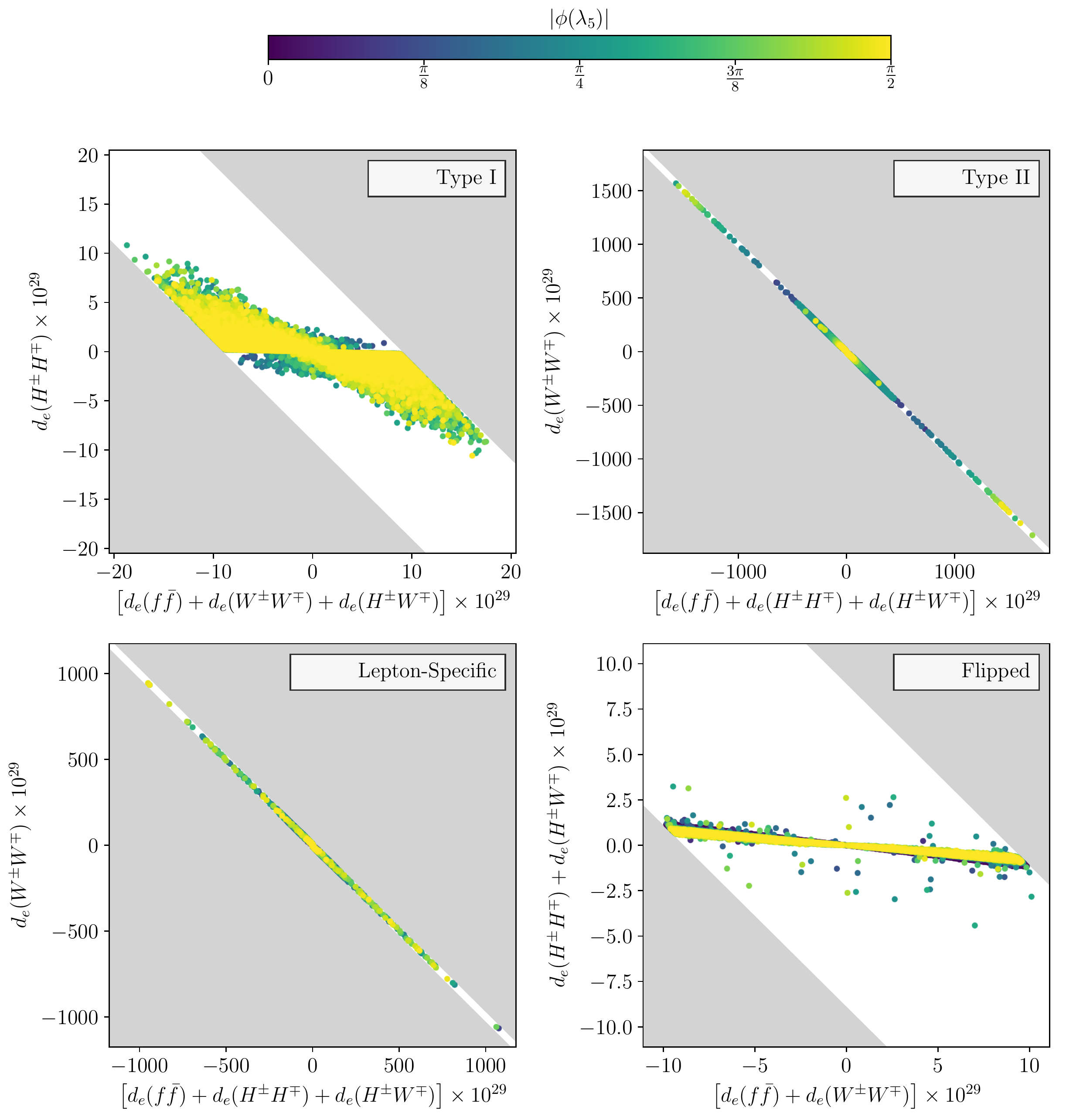}
  \caption{Contributions to the EDM according to their relative sign in Type I (top left),
  Type II (top right), Lepton-Specific (bottom left) and Flipped (bottom right).
  The colour code represents the absolute value of the magnitude of $\phi (\lambda_5)$.
  The grey shaded regions represent the parameter space excluded by the EDM constraints only. The coloured
  points are the ones that passed all the constraints, that is, EDMs plus theoretical and all other
  experimental constraints.} \label{fig:EDMs}
\end{figure}
In figure~\ref{fig:EDMs} we show the individual contributions to the EDM coming from $W$-loops, fermion-loops, charged Higgs loops and charged Higgs plus $W$-loops.
For each C2HDM type, we have grouped the contributions to the EDM according to their relative sign.
For example, in Type II the contributions of the $W$-loops (y-axis) and the sum of the contributions of the fermion loops, charged Higgs loops and charged Higgs plus $W$-loops (x-axis) have opposite sign.
The grey shaded region represents the parameter space excluded by the EDM constraints only.
The colour code represents the absolute value of the magnitude of $\phi (\lambda_5)$.
The first important difference between the models is that the maximum value of the individual contributions is around two orders of magnitude smaller in Type I than in Type II.
This implies that Type I is less constrained by the EDM bounds.
Therefore the remaining constraints play a much more important role in Type I than in Type II.
This also leads to the distribution of values for the CP-violating phase in the figure. In Type II large values of $\phi (\lambda_5)$ prefer regions where either the EDM contributions are very small, {\it i.e.} there are cancellations between loops in the individual contributions, or rely on huge cancellations between different contributions. This is in contrast to Type I, where large values of the CP-violating phase can be found all over the allowed region.
The Flipped case behaves roughly like Type I, while the
Lepton-Specific case behaves like Type II. This is very different from
the usual behaviour of the Yukawa types. The reason is that most
observables are dominated by quark effects giving similar behaviour to
Type I/Lepton-Specific and Type II/Flipped. Since we
are, however, looking at the EDM of the electron the lepton Yukawa
couplings are the most important ones and those are equal in Type I/Flipped and Type II/Lepton-Specific.

Finally, it is important to comment on the different impact of the EDM constraints on the different models, regarding the possibility of having large pseudoscalar Yukawa couplings. For simplicity, we focus our discussion on the $H_1=h_{125}$ case.
In Type I the pseudoscalar Yukawa couplings $c^o_f$ are proportional to $\sin (\alpha_2)/\tan \beta$
for any fermion type $f$. Since $\alpha_2$ is at most of the order $20^\circ$ and $\tan \beta$ is constrained to be above 1, $c^o_f$ will be below about 0.4.
For the other Yukawa types either $c^o_b$ or $c^o_\tau$ are proportional to $\sin (\alpha_2) \tan \beta$. In this case, even if $\alpha_2$ is small the pseudoscalar coupling can be substantial for large $\tan \beta$. We have observed that in Type II the values of $\alpha_2$ can go up to $20^\circ$ while in the Flipped model they barely reach $5^\circ$. In both cases, $c_b^o$ could still be large because large values of $\tan \beta$ are allowed in both models. So what is the reason for having large $c^o_b$ for the Flipped model but not for Type II when $H_1= h_{125}$?
The main reason is that the EDM constraints are less stringent in the Flipped model than in Type II due to relative signs between the CP-odd lepton Yukawa couplings and the remaining CP-odd Yukawa couplings. This ends up flipping several signs in one model relative to the other leading to cancellations between loops and much smaller individual EDM contributions.
In Type II we found that this leads to the result that we can have large $\tan \beta$ only when $\alpha_2$ is very close to zero which is not the case for the Flipped model.

\section{Higgs-to-Higgs decays}
\label{sec:higgstohiggs}

In the last section we have discussed the relation between some of the classes of decays
that probe CP-violation with a number of variables proposed in the literature to
measure CP-violation in the scalar sector. These particular classes were chosen not
only because they can yield large production rates but also because there are currently
searches being performed for these channels at the LHC that have already started during Run 1.
There are, however, other classes of decays that can probe CP-violation and were the subject of a study
that led to the production of benchmarks for Run 2 \cite{Fontes:2015xva, deFlorian:2016spz}.
These other classes involve Higgs-to-Higgs decays such as $H_i \to h_{125} h_{125}$,
$H_j \to H_i h_{125}$ and $h_{125} \to H_i H_i$. In this section we will study what
is the role of the Higgs-to-Higgs decays in the search of CP-violation in the combination
of three decays. The combination of the decays
\begin{equation}
H_{\downarrow \uparrow} \to h_{125} h_{125}, \quad H_{\downarrow \uparrow}  \to h_{125} Z, \quad h_{125} \to ZZ\, ,
\end{equation}
or also
\begin{equation}
h_{125} \to H_{\downarrow \uparrow} H_{\downarrow \uparrow}, \quad h_{125}
\to H_{\downarrow \uparrow} Z, \quad H_{\downarrow \uparrow} \to ZZ\, ,
\end{equation}
are a clear sign of CP-violation, and include Higgs-to-Higgs processes.
Other classes like for instance
\begin{equation}
H_\uparrow \to h_{125} h_{125}, \quad H_\downarrow  \to h_{125} h_{125}, \quad h_{125} \to ZZ \, .
\end{equation}
are not possible in a CP-conserving 2HDM but are possible in the
C2HDM. They are not,
however, a sign of CP-violation. In fact, any model with three CP-even scalars can
have this particular combination of three decays.

\begin{figure}[t]
  \centering
  \includegraphics[width=0.9\linewidth]{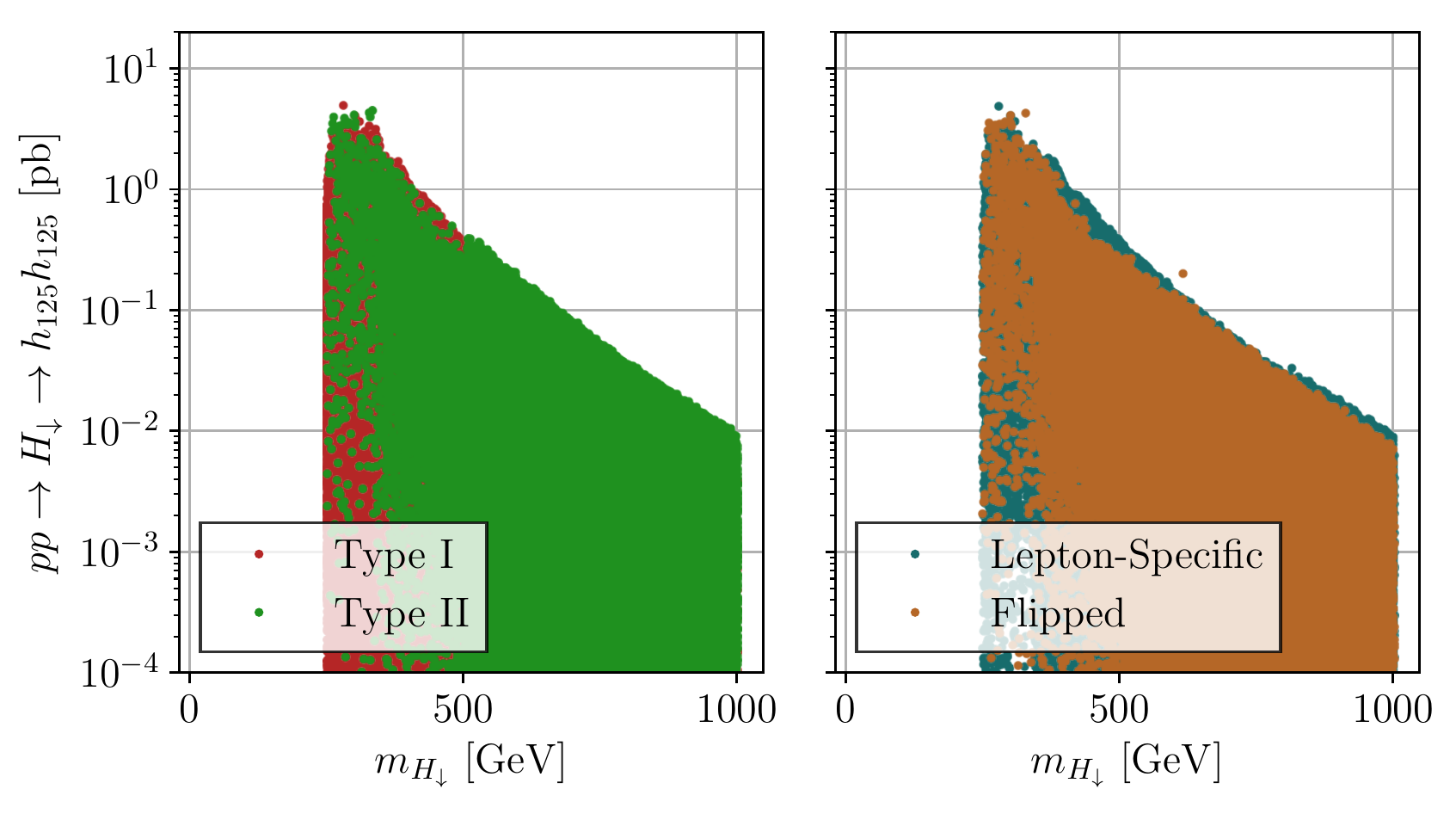}
  \includegraphics[width=0.9\linewidth]{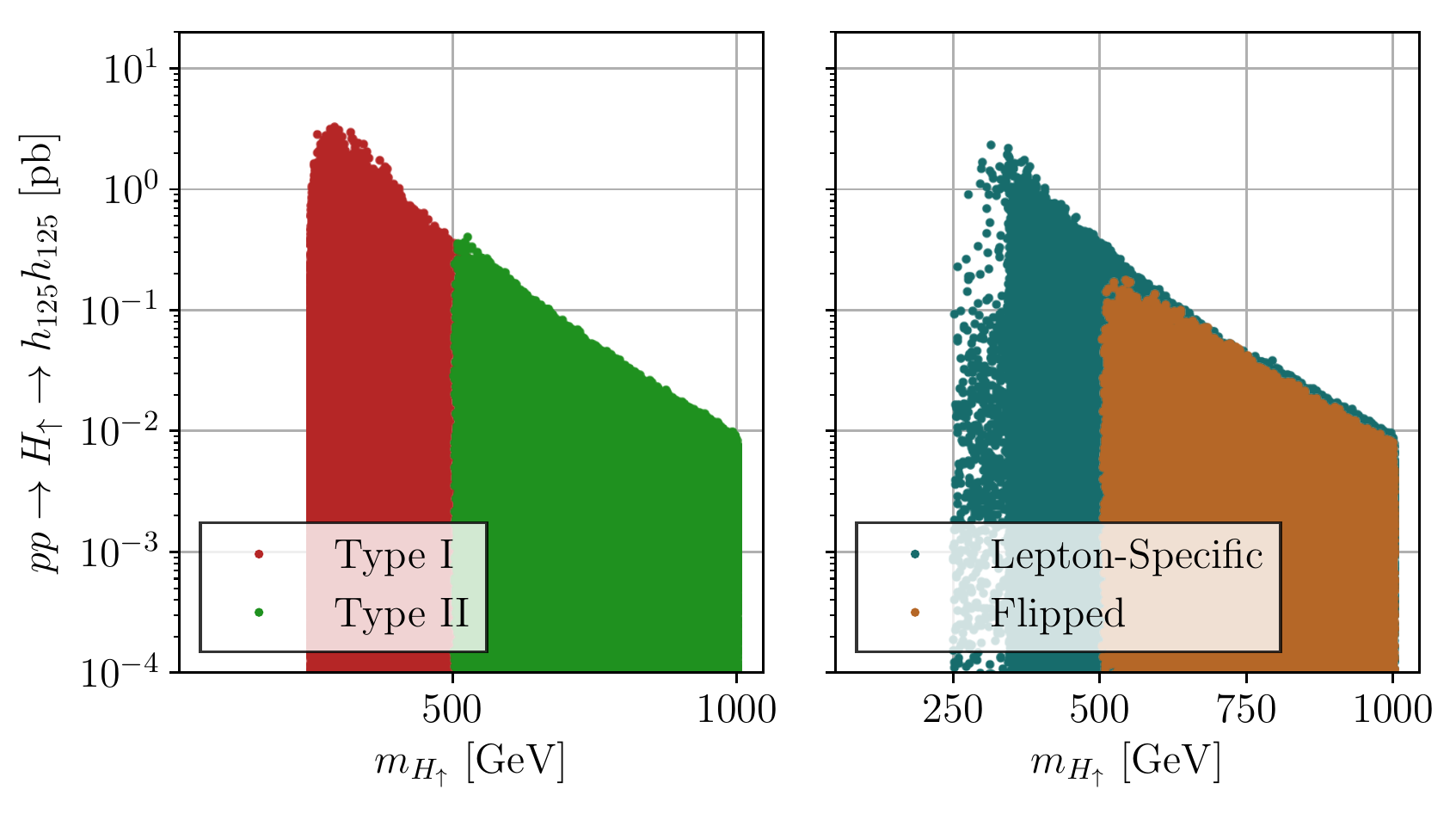}
  \caption{Production rates for the processes
$pp \to H_\downarrow \to h_{125} h_{125}$ (top row) and
  $pp \to H_\uparrow \to h_{125} h_{125}$ (bottom row)
as a function of the respective mass for all C2HDM types.} \label{fig:HHsmHsm}
\end{figure}
\begin{figure}[t]
  \centering
  \includegraphics[width=0.9\linewidth]{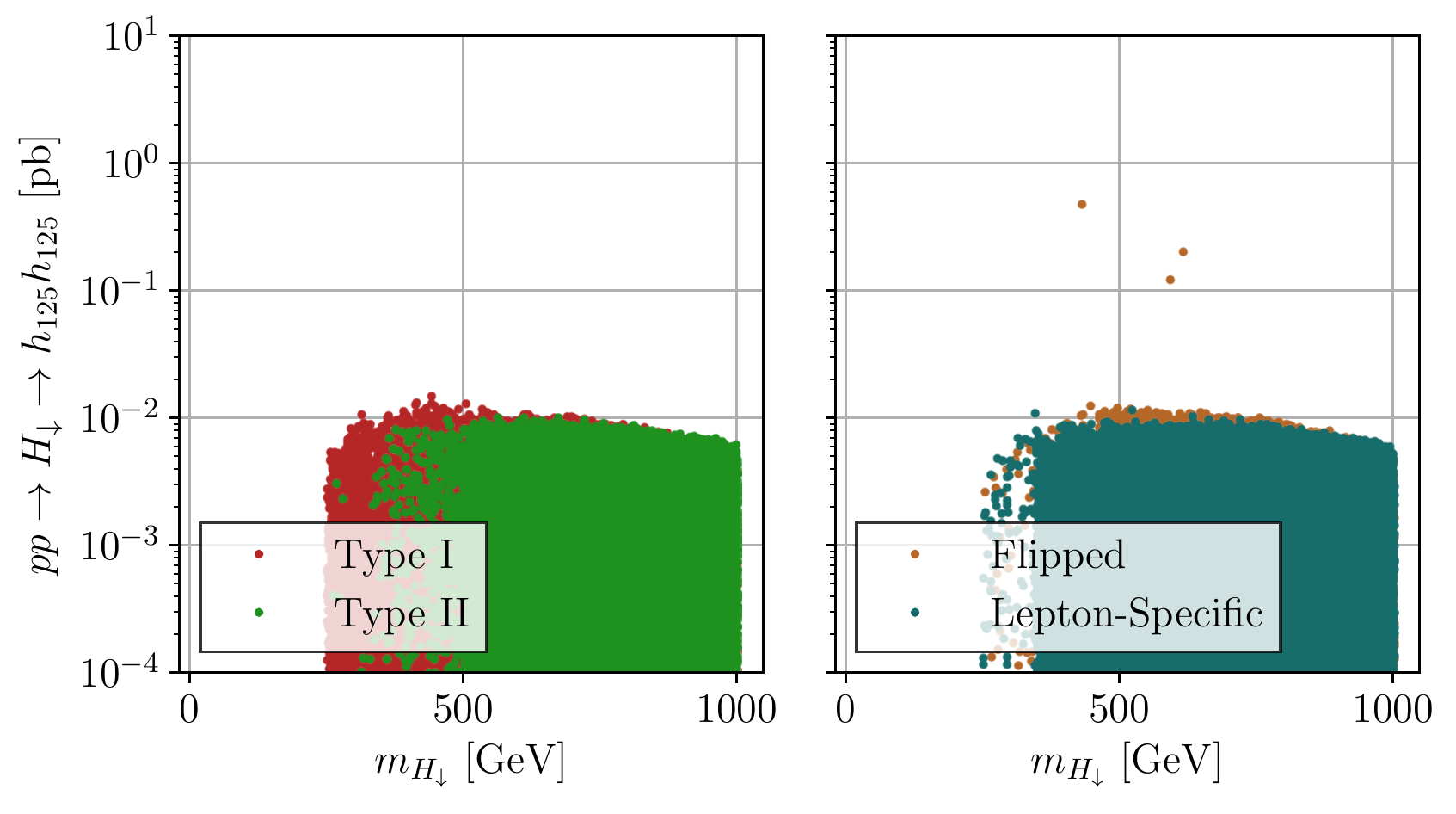}
  \includegraphics[width=0.9\linewidth]{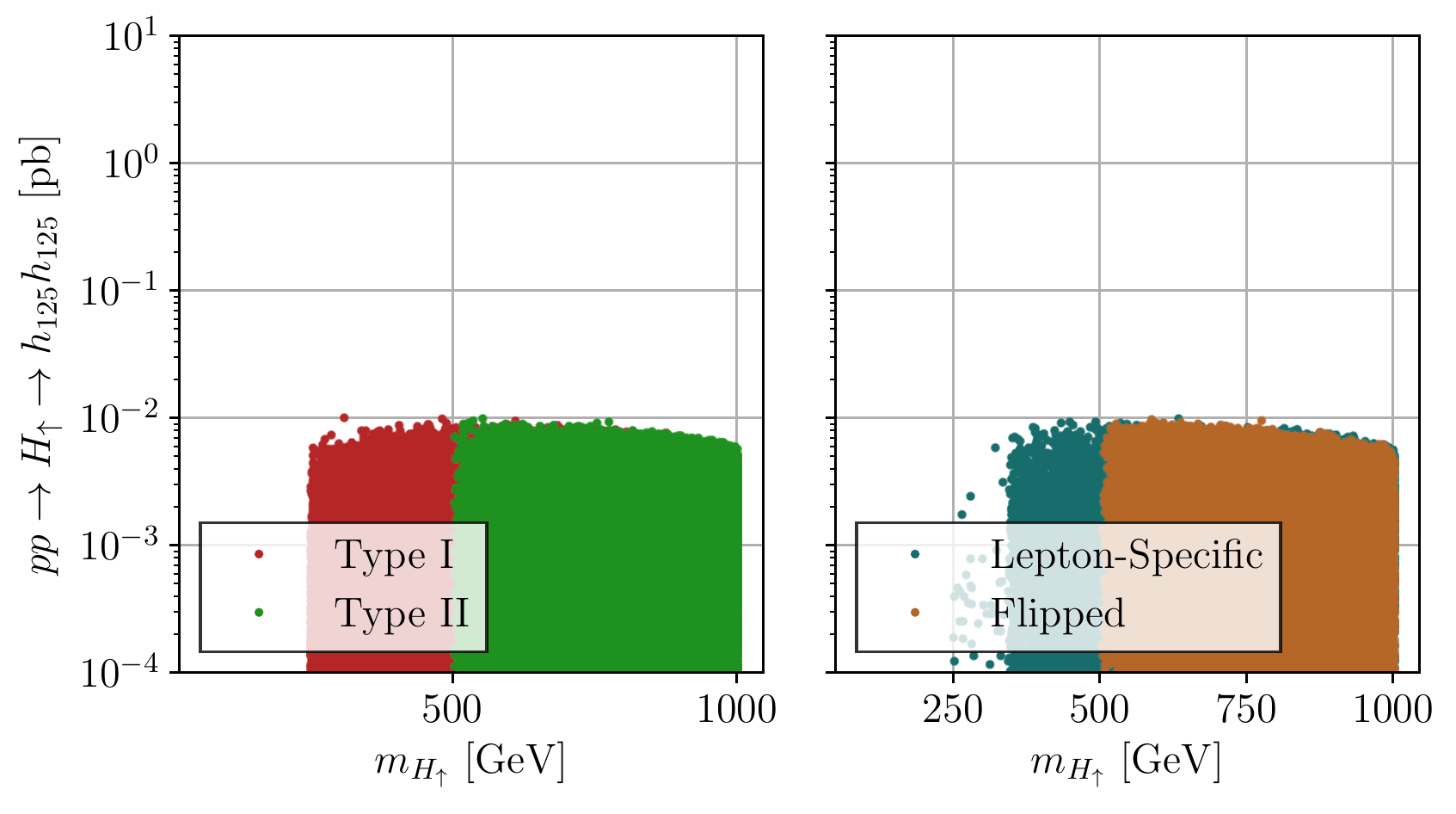}
  \caption{Production rates for the processes
$pp \to H_\downarrow \to h_{125} h_{125}$ (top row) and
$pp \to H_\uparrow \to h_{125} h_{125}$ (bottom row) as a
function of the respective mass for the four C2HDM types (same as figure~\ref{fig:HHsmHsm})
but with the extra condition
  $\sigma (pp \to H_\downarrow \to ZZ) < 1 \,
  \textrm{fb}$ for the top plots and $\sigma (pp \to H_\uparrow \to ZZ) < 1 \, \textrm{fb}$
  for the bottom plots.} \label{fig:HHsmHsmlow}
\end{figure}
In figure~\ref{fig:HHsmHsm} we present the production rates for the processes $pp \to H_\downarrow \to h_{125} h_{125}$ (top row) and   $pp \to H_\uparrow \to h_{125} h_{125}$ (bottom row) as a function of the respective mass for all C2HDM types. In all cases, the rates decrease as one increases the decaying scalar mass. In the four types, the  $pp \to H_\downarrow \to h_{125} h_{125}$
rates can be quite large, reaching about 4 pb in all types.
The maximum values are similar in Type I and Lepton-Specific for  $pp \to H_\uparrow \to h_{125} h_{125}$.
In contrast, for Type II and Flipped, the largest rates in $pp \to
H_\uparrow \to h_{125} h_{125}$ decrease by about an order of
magnitude because in these cases the heavier neutral scalar cannot be
much lighter than the charged Higgs boson, which is heavy to comply with $B$-physics constraints.
In order to understand how relevant the searches for the two scalar final states   are we show in figure~\ref{fig:HHsmHsmlow} the same rates as in the previous figure~\ref{fig:HHsmHsm} but with the extra condition   $\sigma (pp \to H_\downarrow \to ZZ) < 1 \, \textrm{fb}$ for the top plots and $ \sigma (pp \to H_\uparrow \to ZZ) < 1 \, \textrm{fb}$
for the lower plots. It is clear from the plots that,   with the extra
restriction on the $ZZ$ final state, the cross sections now barely
reach 10 fb for the two decay scenarios and for all types.
Hence, although possible, it will be very hard to detect the new scalars in the $h_{125} h_{125}$ final state if they are not detected in the $ZZ$ final state. One should note that the cross section for di-Higgs production in the SM
is about 33 fb. Consequently, a resonant di-Higgs final state such as the one presented
in figure~\ref{fig:HHsmHsm} would easily be detected because the cross sections can reach the pb level.
However, it is also clear that once we force $ \sigma (pp \to H_\uparrow \to ZZ) < 1 \, \textrm{fb}$ it is no longer possible to detect these di-Higgs states even at the High Luminosity LHC.

\begin{figure}[t]
  \centering
  \includegraphics[width=0.9\linewidth]{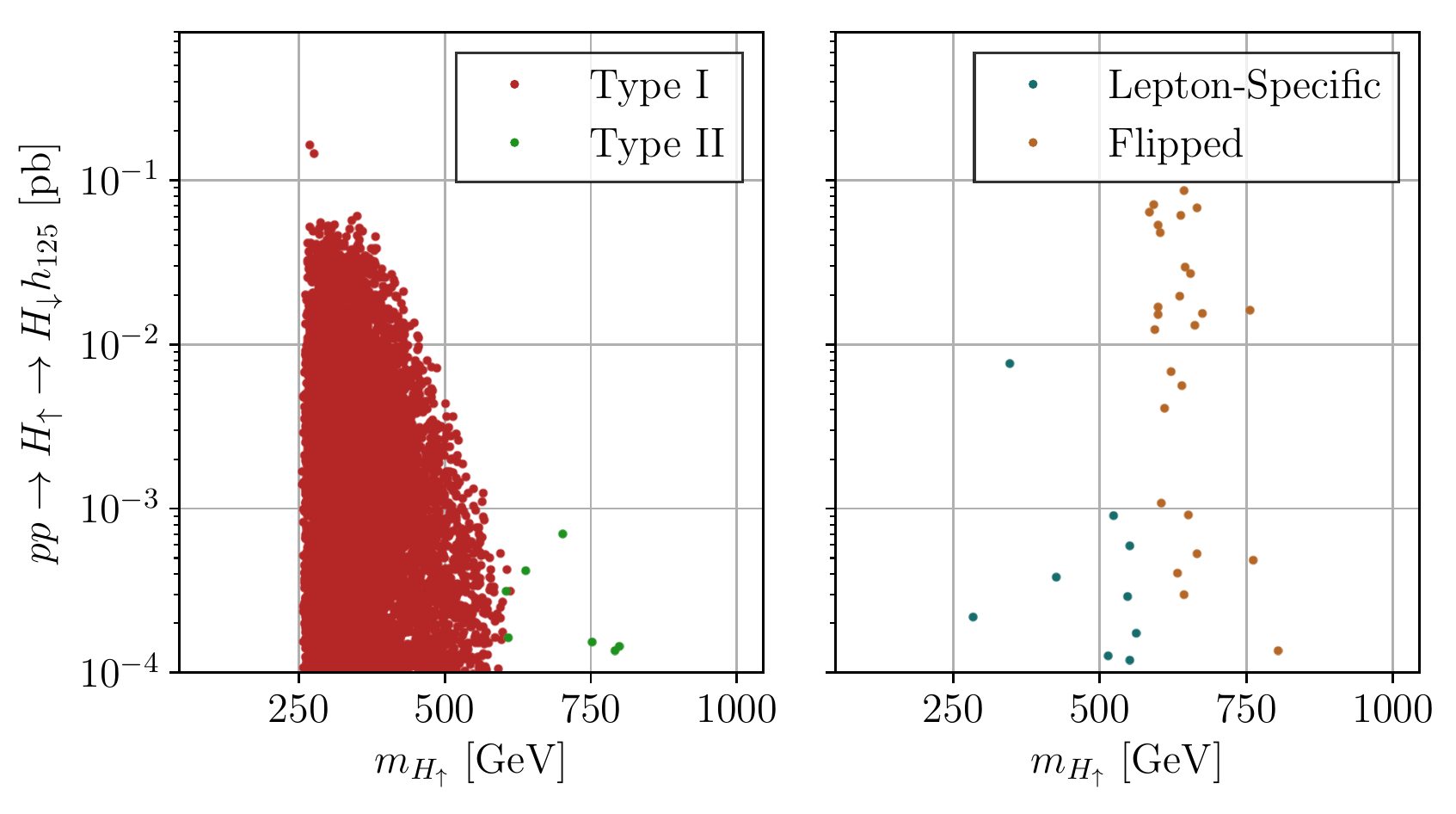}
  \caption{Production rates for the process $pp \to H_\uparrow \to H_\downarrow h_{125}$ as a function
  of the heavier Higgs boson mass, for all C2HDM types.} \label{fig:HHlHsm}
\end{figure}

\begin{figure}[t]
  \centering
  \includegraphics[width=0.9\linewidth]{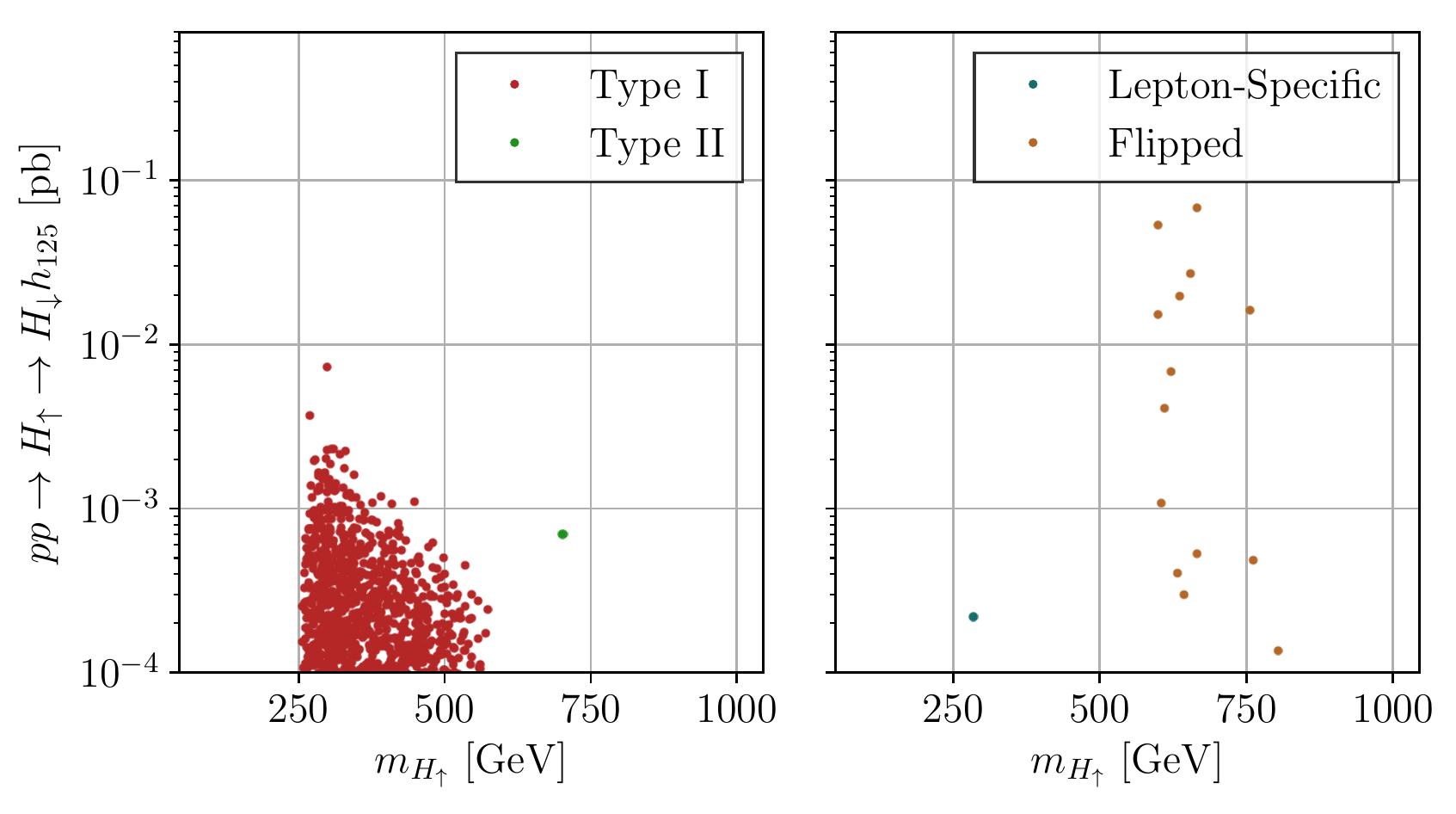}
  \caption{Production rates for the process $pp \to H_\uparrow \to H_\downarrow h_{125}$ as a function of
  the heavier Higgs boson mass, for all C2HDM types (same as figure~\ref{fig:HHlHsm}) but with the extra condition
  $\sigma (pp \to H_\uparrow \to ZZ) < 1 \, \textrm{fb}$.} \label{fig:HHlHsmlow}
\end{figure}

In figure~\ref{fig:HHlHsm} we show the production rates for the process
$pp \to H_\uparrow \to H_\downarrow h_{125}$ as a function
  of the heavier Higgs mass, for all C2HDM types.
For this channel the rates can reach at most about 100 fb,
and only for Type I and Flipped.
In Type II the rates are at most at the fb level.
The  rates for the
  $H_\downarrow h_{125}$ final state with the extra condition
$ \sigma (pp \to H_\uparrow \to ZZ) < 1 \, \textrm{fb}$
are shown in figure~\ref{fig:HHlHsmlow}.
The maximum rates (for low masses) are now reduced by about a factor of 5
for Type I.
However, the rates do not decrease much for the Flipped C2HDM,
and some signal at LHC Run 2 could point to this C2HDM type.
  Finally, although $H_\uparrow \to H_\downarrow h_{125}$ appears hard to detect in these models
  it is nevertheless a clear signal of non-minimal models and should therefore be a priority for the LHC Run 2.

\begin{figure}[t]
  \centering
  \includegraphics[width=0.9\linewidth]{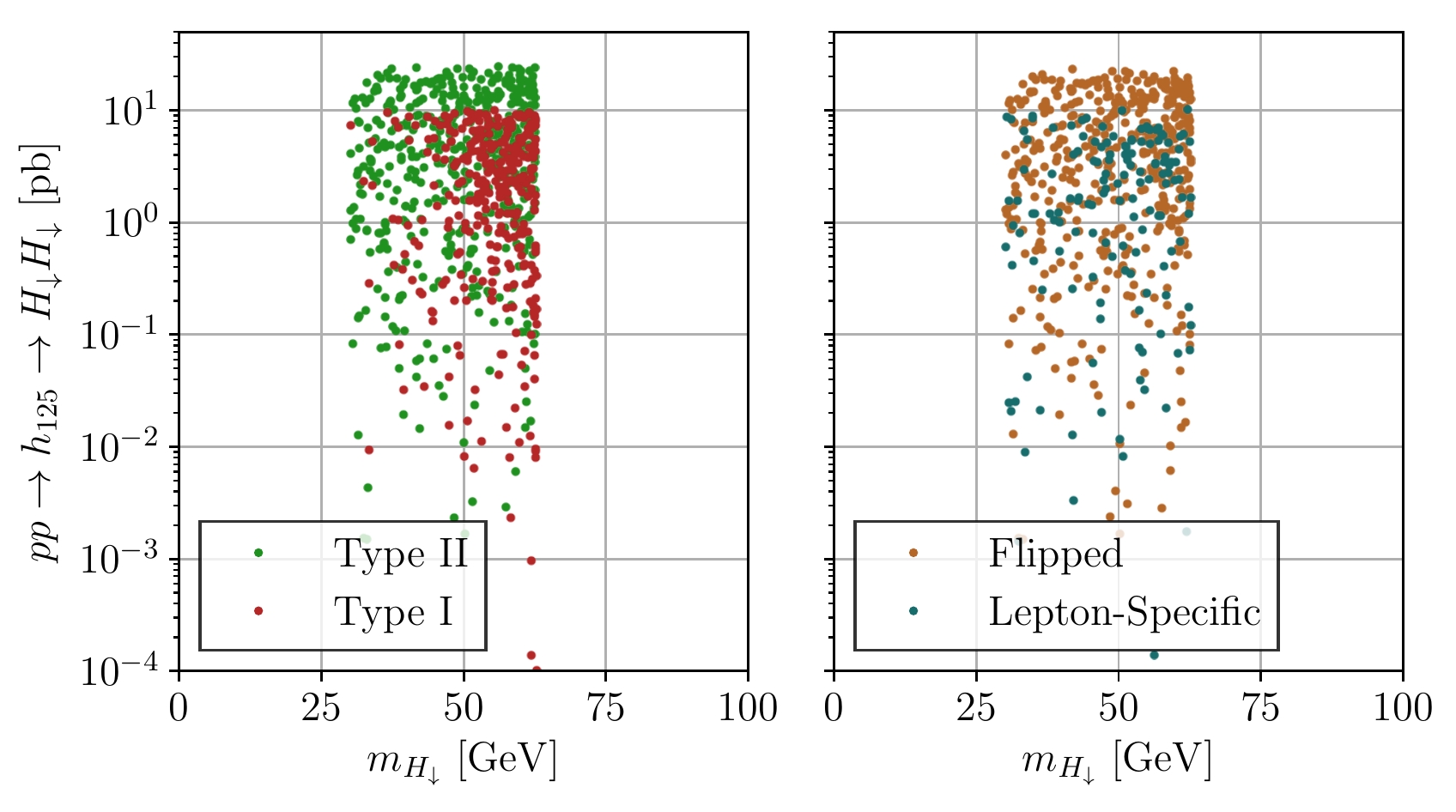}
  \caption{Production rates for the process
	$pp \to h_{125} \to H_\downarrow H_\downarrow $ as a function
  of the lighter Higgs boson mass.} \label{fig:HsmTOHlHl}
\end{figure}
We end this section with the production rate for the process
$pp \to h_{125} \to H_\downarrow H_\downarrow $ as a function
of the lighter Higgs mass for the various C2HDM types,
which are shown in figure~\ref{fig:HsmTOHlHl}.
Most points correspond to a mass of the heavier state above 125 GeV.
But, as shown in figure~\ref{fig:rates_HsmEqH3_HlHl},
in the Type I and Lepton-Specific cases
there are still solutions with $H_3 = h_{125}$.
Here the rates can be quite large if
the lightest Higgs has a mass below 60 GeV.
For this region the production rates can reach 10 pb (30 pb)
for Type I and Lepton-Specific (for Type II and Flipped).
In order to understand what would be the behaviour when choosing
a definite final state, we have checked that the
rates $pp \to h_{125} \to H_\downarrow H_\downarrow \to b \bar b \tau^+ \tau^-$
are still above the pb level for all model types.

\section{Conclusions}
\label{sec:concl}

In this work we have analysed in detail a minimal complex version of the 2HDM, known as the C2HDM.
The inclusion of all available experimental and theoretical constraints allowed us to present
an up to date status of the model. We have shown that large CP-odd Yukawa couplings of $h_{125}$ are still possible in all Yukawa types except Type I. However, in Type II this is only a possibility if $H_2=h_{125}$. We provided two different interesting kinds of benchmark points which are in agreement with all current observations: for $h_{125}$ coupling like a scalar to some fermions and like a pseudoscalar to others, and for scenarios where the scalar and pseudoscalar Yukawa couplings of $h_{125}$ to some fermion are of similar size.

The model has only one CP-violating parameter. In previous works we have presented classes
of three decays that are a clear sign of CP-violation, when all three are observed. In this work
we looked for correlations between the production rates of the processes in each class and
the phase $\phi (\lambda_5)$ that measures CP-violation. We have shown that there is no correlation
for the most relevant classes. We have then tested the correlation with other
CP-violating variables proposed in the literature. The conclusion is that
some correlation can be seen between the production rates and the variables. This
is particularly true for the variable $\xi_V$ and even more for Type II.
However, in most cases there is almost no correlation between the high rates
of the CP-violating processes and the proposed CP-violating variables.
The results also tell us that measuring small rates should not be interpreted
as a sign of a small CP-violating angle.

Finally, we presented in the last section the production rates for the scenarios
where the Higgs bosons decay to two other scalars. The search for a scalar decaying
into two $h_{125}$ Higgs bosons was already performed during Run 1. However,
the search for $H_i \to H_j h_{125}$ has not started yet. It is clear from the results
that it will be much harder to probe CP-violation with classes of decays that
involve scalar to scalar decays. However, on one hand it could be that all
other decays would be very constrained. On the other hand, the observation
of scalar to scalar decays would be a first step to reconstruct the Higgs potential.
Hence, these searches should be a priority for Run 2 and especially for the high luminosity LHC.

As part of a common effort to a proper interpretation of the LHC results we release
the code {\tt C2HDM\_HDECAY} that calculates the decay widths and branching ratios of all the C2HDM scalars including state-of-the-art QCD corrections.

\section*{Appendix}
\setcounter{equation}{0}
\begin{appendix}

\section{Feynman rules for the C2HDM  \label{app:c2hdm_FR}}

We collect here the couplings of the neutral and charged Higgs bosons in
the C2HDM in the unitary gauge. The conventions are that all particles
and momenta are incoming into the vertex. As for the SM subset
we use the notation for the covariant derivatives
contained in \cite{Romao:2012pq},
with all $\eta$'s positive.
The complete set of Feynman rules for the C2HDM \cite{C2HDM_FR} may
be found at the url: \\
\centerline{\tt http://porthos.tecnico.ulisboa.pt/arXiv/C2HDM/}

\subsection{Couplings of neutral Higgs bosons to fermions}

The couplings of neutral Higgs bosons to fermions can be written in general for all the
neutral Higgs bosons of the model $H_i$ in a compact form
\begin{equation}
\label{eq:1}
  \mathcal{L}= - \frac{m_f}{v}\ \overline{f}
\left[c^e(H_i ff) +i \gamma_5\, c^o (H_i ff)   \right] f\, H_i,
\end{equation}
with the coefficients presented in table~\ref{tab:yukcoup}.

\subsection{Couplings of charged Higgs bosons to fermions}

The couplings of the charged Higgs bosons to fermions can be expressed in the
following Lagrangian
\begin{equation}
  \label{eq:2}
  \mathcal{L}= \frac{\sqrt{2}}{v}\, \overline{\psi}_{d_{i}} \left[\vb{12}
    m_{\psi_{d_{i}}} \eta_L P_L
+ m_{\psi_{u_{i}}} \eta_R P_R \right] \psi_{u_{i}} H^{-}
  + \frac{\sqrt{2}}{v}\, \overline{\psi}_{u_{i}} \left[\vb{12}
    m_{\psi_{d_{i}}} \eta_L P_R + m_{\psi_{u_{i}}} \eta_R P_L \right] \psi_{d_{i}} H^{+},
\end{equation}
where $i$ are generation indices, $(\psi_{u_{i}},\psi_{u_{i}})=(u_i,d_i), (\nu_i,\ell_i)$, for
quarks and leptons in an obvious notation and $P_L = (1-\gamma_5)/2$
and $P_R = (1+\gamma_5)/2$. The couplings
$\eta_{L,R}$ are given in table~\ref{tab:2}.
In these expressions,
we neglect the masses of the neutrinos, so in the last line in
table~\ref{tab:2} the zeros mean that the corresponding mass in
eq.~(\ref{eq:2}) is zero.
\begin{table}[htb]
  \centering
  \begin{tabular}{|l|c|c|c|c|}\hline
    & Type I & Type II & Lepton & Flipped\\
    &  &  & Specific & \\\hline
$\eta^q_L$ &$ -\cot\beta$ & $\tan\beta$ & $ -\cot\beta$ &$\tan\beta$ \\
$\eta^q_R$ &$\cot\beta$ &$\cot\beta$  &$\cot\beta$  & $\cot\beta$ \\
$\eta^\ell_L$ &$ -\cot\beta$ & $\tan\beta$ & $\tan\beta$ & $ -\cot\beta$\\
$\eta^\ell_R$ &0 & 0 & 0 &0 \\\hline
      \end{tabular}
  \caption{Couplings of the charged Higgs bosons to fermions.}
  \label{tab:2}
\end{table}

\subsection{Cubic interactions of neutral Higgs bosons}

The cubic interactions of neutral Higgs bosons, $[H_i,H_j,H_k]$, have long
expressions. We do not write them here but we collect them in one web
page~\cite{C2HDM_FR}.
The
expressions there are the Feynman rules without the $i$.

\subsection{Cubic interactions of neutral and charged Higgs bosons}

These are (on the right-hand side of these expressions we write the
Feynman rule including the $i$),

\begin{align}
[H_i, H^+, H^{-}]&=
-i\, v \left[ \text{Im}(\lambda_5) R_{i3} \cos(\beta) \sin(\beta)
\right. \nonumber\\[+2mm]
&\hskip 15mm
+
  R_{i1}  \cos(\beta) (\lambda_3
      \cos(\beta)^2
- (\text{Re}(\lambda_5) - \lambda_1 +
       \lambda_4) \sin(\beta)^2)  \nonumber\\[+2mm]
&\hskip 15mm
\left.+
  R_{i2}  (-(\text{Re}(\lambda_5) - \lambda_2
+ \lambda_4) \cos(\beta)^2 \sin(\beta)+
    \lambda_3 \sin(\beta)^3)\right]\label{eq:h1HpHm}\nonumber\\[+2mm]
&\equiv i\, \lambda_i \, v \equiv i\, g_{H_{i}H^+H^-}\, ,
\end{align}
where the $\lambda_i$ or $g_{H_{i}H^+H^-}$can be read from
eq.~(\ref{eq:h1HpHm}). The $\lambda_i$ are in the
notation used in ref.~\cite{Fontes:2014xva} and should not be confused
with the parameters in the potential.

\subsection{Cubic interactions with gauge bosons}

\textit{One gauge boson}
\begin{align}
[H_i, H^{\mp}, W^{\pm\, \mu}]&=\pm i\, \frac{g}{2} \left( p_1-p_{\mp}\right)^\mu
\left(R_{i1} \sin(\beta) - R_{i2} \cos(\beta) \mp i\, R_{i3}
\right)\label{eq:h1HmWp}\, ,
\\[+2mm]
[H_i, H_j, Z^{\mu}] &=\frac{g}{2 c_W} \left(p_i-p_j\right)^\mu
  \epsilon_{ijk} \left[
    \cos(\beta) R_{k1} + \sin(\beta) R_{k2} \right]
\label{eq:h2h3Z}\, ,
\\[+2mm]
[A^{\mu}, H^+, H^{-}]&= -i\; e \left( p_+-p_{-}\right)^\mu\, ,
\\[+2mm]
[H^+, H^{-}, Z^{\mu}]&=
-i\; \frac{g}{2 c_W} (c_W^2-s_W^2) \left( p_+-p_{-}\right)^\mu\, .
\end{align}
\textit{Two gauge bosons}
\begin{align}
[H_i, W^+_{\mu}, W^{-}_{\nu}]&=i\, g M_W\, g_{\mu\nu}
 \left[R_{i1} \cos(\beta) +  R_{i2} \sin(\beta)\right] \equiv i\, g
 M_W\, g_{\mu\nu} \ C_i\label{eq:h1WpWm}\, ,
\\[+2mm]
[H_i, Z_{\mu}, Z_{\nu}]&=i\, \frac{g M_Z}{c_W}\, g_{\mu\nu}
 \left[R_{i1} \cos(\beta) +  R_{i2} \sin(\beta)\right]
\equiv i\, \frac{g M_Z}{c_W}\, g_{\mu\nu} \ C_i\label{eq:h1ZZ}\, ,
\end{align}
where in eq.~(\ref{eq:h1WpWm}) and
eq.~(\ref{eq:h1ZZ}) we used a notation to
make contact with our previous conventions~\cite{Fontes:2014xva}.

\subsection{Quartic interactions with Higgs bosons}

Again these interactions involve quite long expressions and they are given in
our web page~\cite{C2HDM_FR}. These include, $[H_i,H_j,H_k,H_l]$
where $i,j,k,l=1,2,3$ and $[H_i,H_j, H^-,H^+]$.

\subsection{Quartic interactions with Gauge bosons}

\begin{align}
[A_{\mu}, A_{\nu}, H^+, H^{-}]&= 2\, i\, e^2\, g_{\mu\nu}\, ,
\\[+2mm]
[A_{\mu}, H_i, H^{\mp}, W^\pm_{\nu}]&=-i\, \frac{e g}{2}
\left[ R_{i1} \sin(\beta)-  R_{i2} \cos(\beta)
 \mp i\, R_{i3} \right]\, g_{\mu\nu}\label{eq:Ah3HmWp1}\, ,
\\[+2mm]
[H^+, H^{-}, W^+_{\mu}, W^{-}_{\nu}]&
=i\, \frac{g^2}{2}\, g_{\mu\nu}\, ,
\\[+2mm]
[H_i, H_j, W^+_{\mu}, W^{-}_{\nu}]&= i\,
\frac{g^2}{2}\,g_{\mu\nu}\, \delta_{ij}\, ,
\\[+2mm]
[A_{\mu}, H^+, H^{-}, Z_{\nu}]&=
i\, \frac{e g}{c_W} (c_W^2-s_W^2) g_{\mu\nu}\, ,
\\[+2mm]
[H_i, H^{\mp}, W^\pm_{\mu}, Z_{\nu}]&=i\,
\frac{e^2}{2 c_W}
\left[ R_{i1} \sin(\beta) -  R_{i2} \cos(\beta) \mp i\, R_{i3}
    \right]\, g_{\mu\nu}\, ,
\\[+2mm]
[H^+, H^{-}, Z_{\mu}, Z_{\nu}]&= i\, \frac{g^2}{2 c_W^2}
(c_W^2-s_W^2)^2 g_{\mu\nu}\, ,
\\[+2mm]
[H_i, H_j, Z_{\mu}, Z_{\nu}]&=i\, \frac{g^2}{2 c_W^2}
g_{\mu\nu}\, \delta_{ij}\, .
\end{align}

\section{The Fortran Code {\tt C2HDM\_HDECAY}  \label{app:c2hdmhdecay}}

The code {\tt C2HDM\_HDECAY} is the implementation of the CP-violating
2HDM in the program {\tt
  HDECAY v6.51}~\cite{Djouadi:1997yw,Butterworth:2010ym}, which is written
in Fortran77.  All changes with respect to the C2HDM have been
included in the main file {\tt hdecay.f}, which is now called {\tt
  chdecay.f}. Further linked routines
have been taken over from the original {\tt HDECAY} program, so that
the code is completely self-contained. The decay widths are computed including the
most important state-of-the-art higher order QCD corrections and the
relevant off-shell decays. Note, that it does not include off-shell
Higgs-to-Higgs decays, but only on-shell decays into a lighter Higgs
pair. The QCD corrections can be taken over from the SM and the
minimal supersymmetric extension (MSSM), respectively, for which {\tt
  HDECAY} was originally designed. The electroweak corrections on the
other hand cannot be adapted from the available corrections in the SM
and/or MSSM so that they have been consistently turned off. \s

The C2HDM input parameters are specified in the input file {\tt
  c2hdecay.in} which has been obtained by extending the original {\tt
  HDECAY} input file {\tt hdecay.in}. The C2HDM branching ratios and
total widths are calculated after setting the input value {\tt
  C2HDM}$=1$ in {\tt c2hdecay.in}. The required input parameters are
set in the block '{\tt complex 2 Higgs Doublet Model}'. Here the user
specifies the values of two of the neutral Higgs boson masses, the
third one is computed from the input values, and the charged Higgs boson
mass. Furthermore, the mixing angles $\alpha_{1,2,3}$ and $\tan\beta$
have to be set as well as the real part of the mass parameter
$M_{12}^2$. The type of the fermion sector is chosen via the input
variable {\tt TYPE\_cp}. The values 1, 2, 3 and 4 correspond to the I,
II, Lepton-Specific and Flipped types, respectively. For illustration,
we display part of an example input file for the C2HDM case.

\vspace*{0.2cm}
\begin{Verbatim}[fontsize=\small,commandchars=\\\{\},xleftmargin=10mm]
C2HDM    = 1
...
*********************** complex 2 Higgs Doublet Model *********************
M1_2HDM  = 125.D0
M2_2HDM  = 4.96226790D2
MCH_2HDM = 4.9242445D2
alp1_2HDM= 0.88941955D0
alp2_2HDM= -0.096916989D0
alp3_2HDM= 1.05235430D0
tbetc2HDM= 1.19671762D0
R_M12_H2 = 1.51744100D3
TYPE\_cp  = 2
**************************************************************************
...
\end{Verbatim}

The code is compiled with the file {\tt makefile}. By typing {\tt
  make} an executable file called {\tt run} is produced. The program
is executed with the command {\tt run}. It calculates the branching ratios
and total widths which are written out together with the mass of the
decaying Higgs boson. The names of the output files are {\tt
  br.Xy\_C2HDM}. Here {\tt X=H1, H2, H3, c} denotes the decaying
Higgs particle, where '$c$' refers to the charged Higgs boson. Files
with the suffix {\tt y=a} contain the branching
ratios into fermions, with {\tt y=b} the ones into gauge bosons and
the ones with {\tt y=c, d} the branching ratios into lighter Higgs
pairs or a Higgs-gauge boson final state. For illustration, we present
the example of an output that has been obtained from the above input file. The
produced output in the four output files {\tt br.H3y\_C2HDM} for the
heaviest neutral Higgs boson is given by

\vspace*{0.2cm}
\begin{verbatim}
   MH3         BB       TAU TAU     MU MU         SS         CC         TT
_______________________________________________________________________________

 506.461     0.1122E-02 0.1600E-03 0.5658E-06 0.4094E-06 0.2180E-04 0.9505
   MH3           GG     GAM GAM     Z GAM         WW         ZZ
_______________________________________________________________________________

 506.461     0.3149E-02 0.1046E-04 0.3603E-05 0.1324E-01 0.6341E-02
     MH3       Z H1       Z H2     W+- H-+
_______________________________________________________________________________

 506.461     0.1807E-01 0.5732E-08 0.1366E-06
      MH3      H1H1       H1H2       H2H2      H+ H-      WIDTH
_______________________________________________________________________________

 506.461     0.7417E-02  0.000      0.000      0.000      10.21
\end{verbatim}

The program files can be downloaded at the url: \\
\centerline{\tt http://www.itp.kit.edu/$\sim$maggie/C2HDM} \\
There one can find a short explanation of the program and
information on updates and modifications of the program. Furthermore,
sample output files are given for a sample input file.


\end{appendix}


\vspace*{0.5cm}
\section*{Acknowledgments}
The work of D.F., J.C.R. and J.P.S. is supported in part
by the Portuguese \textit{Funda\c{c}\~{a}o para a Ci\^{e}ncia e Tecnologia}
(FCT) under contracts CERN/FIS-NUC/0010/2015 and UID/FIS/00777/2013.
MM acknowledges financial support from the DFG project “Precision Calculations
in the Higgs Sector - Paving the Way to the New Physics Landscape” (ID: MU 3138/1-1).
\vspace*{0.5cm}

\vspace*{1cm}
\bibliographystyle{JHEP}
\bibliography{c2hdmpaper}
\end{document}